\documentclass[preprint,nofootinbib,superscriptaddress]{revtex4}
\usepackage{amsmath} 
\usepackage{mathrsfs}
\usepackage{amssymb}

\def\be{\begin{equation}}
\def\ee{\end{equation}}
\def\ba{\begin{eqnarray}}
\def\ea{\end{eqnarray}}	
\def\l{\left}
\def\r{\right}
\def\fr{\frac}
\def\la{\label}
\def\d{\partial}
\def\vphi{\varphi}

\begin{document}

\title{Effective Theory Approach to the Spontaneous Breakdown of Lorentz Invariance}

\date{\today}

\author{Cristian Armendariz-Picon}
\affiliation{Department of Physics, Syracuse University, Syracuse, NY 13244-1130, USA}

\author{Alberto Diez-Tejedor}
\affiliation{Instituto de Ciencias Nucleares, Universidad Nacional Aut\'onoma de M\'exico, M\'exico D.F. 04510, M\'exico}

\author{Riccardo Penco}
\affiliation{Department of Physics, Syracuse University, Syracuse, NY 13244-1130, USA}

\title{Effective Theory Approach to the Spontaneous Breakdown of Lorentz Invariance}

\begin{abstract}
We generalize the coset construction of Callan, Coleman, Wess and Zumino to theories in which the Lorentz group is spontaneously broken down to one of its subgroups. This allows us to write down the most general low-energy effective Lagrangian in which Lorentz invariance is non-linearly realized, and to explore the consequences of broken Lorentz symmetry without having to make any assumptions about the mechanism that triggers the breaking.  We carry out the construction both in flat space, in which the Lorentz group is a global spacetime symmetry, and in a generally covariant theory, in which the Lorentz group can be treated as a local internal symmetry.  As an illustration of this formalism,  we construct the most general effective field theory in which the rotation group remains unbroken,  and  show that the latter is just the Einstein-aether theory.

\end{abstract}

\maketitle
\section{Introduction}

It is hard to overemphasize the central role that the Lorentz group plays in our present understanding of nature. The standard model of particle physics, for instance, consists of all renormalizable interactions invariant under Lorentz transformations and its internal symmetry gauge group, which act on the matter fields of the theory. While most standard model  extensions alter either its field content or gauge group, they rarely drop Lorentz invariance. Of course, such  a reluctance has a well-established observational support. Elementary particles appear in (irreducible) representations of the Lorentz group, and their interactions seem to be well described by Lorentz-covariant laws. Lorentz-breaking operators in the standard model of particle physics were first considered by Colladay and Kostelecky  \cite{Colladay:1998fq}, and Coleman and Glashow \cite{Coleman:1998ti}. Experimental and observational constraints on such operators are so stringent  \cite{Kostelecky:2008ts} that it is safe to assume that any violation of Lorentz invariance in the standard model must be extremely small.

The status of the Lorentz group in theories of gravity is somewhat different. Because the group of diffeomorphisms does not admit spinor representations, in generally covariant theories the Lorentz group is introduced as  a \emph{local} \emph{internal} symmetry. Thus, in gravitational theories one formally deals with two distinct groups of transformations: diffeomorphisms and local Lorentz transformations. Even in the context of generally covariant theories, it is thus natural to ask and inquire whether the gravitational interactions respect Lorentz invariance, and what constraints we can impose on any Lorentz-violating gravitational interactions. To date, experimental bounds still allow significant deviations from Lorentz invariance in gravitational interactions \cite{Kostelecky:2003fs, Kostelecky:2008ts,Will:2005va}.

In this article we mainly explore some consequences of broken Lorentz invariance in generally covariant theories. We have in mind here theories that admit  a generally covariant, but not a Lorentz invariant formulation. Generically, the breaking of diffeomorphism invariance in non-trivial backgrounds also leads to Lorentz symmetry breaking \cite{ArkaniHamed:2004ar,  Horava:2009uw, Blas:2009yd}, but the nature of the breaking in these cases is quite different from what we explore here, and indeed leads to different phenomenology.  The spontaneous breaking of Lorentz symmetry (in our sense) has been mostly explored by means of  particular models in which vector fields \cite{Kostelecky:1989jp,Kostelecky:1989jw,Seifert:2009gi,Jacobson:2000xp,Gripaios:2004ms,Libanov:2005vu,Graesser:2005bg} or higher-rank tensors \cite{Altschul:2009ae} develop a non-vanishing vacuum expectation value. In this article we follow a general approach and address consequences that merely follow from  the symmetry breaking pattern, regardless of any specific model of Lorentz symmetry breaking. Such a model-independent approach was first introduced by Weinberg to describe the breakdown of chiral invariance in the strong interactions \cite{Weinberg:1968de}, and was subsequently generalized by Callan, Coleman, Wess and Zumino  to the breaking of any internal symmetry group down to any of its subgroups \cite{Coleman:1969sm,Callan:1969sn}.  Their approach was further broadened to the case of spontaneous breaking of space-time symmetries \cite{Isham:1970gz,Volkov:1973vd,Ogievetsky, Borisov:1974bn} down to the Poincar\'{e} group.  Here, we extend all these results to the case in which  the Lorentz group itself is broken down to one of its subgroups. A naive application of Goldstone's theorem then implies the existence of massless Goldstone bosons, which may in principle participate in long-ranged interactions and alter the Newtonian and post-Newtonian limits of the theory. Equivalently, we may also think of these additional fields as additional polarizations of the graviton.

These considerations are not a purely academic exercise. Motivated by cosmic acceleration, several authors have devoted substantial attention to massive theories of gravity \cite{ArkaniHamed:2002sp,Dubovsky:2004sg,Rubakov:2008nh,Chamseddine:2010ub} and other modifications \cite{Dvali:2000hr,ArkaniHamed:2003uy,Carroll:2004de}, even though the distinction between modifications of gravity and theories with additional matter fields is often blurry. Within the last class, several groups have studied the cosmological dynamics induced by vector fields with non-zero expectation values (see for instance \cite{Ford:1989me, Bento:1992wy,Carroll:2004ai,Jimenez:2008au, Golovnev:2008cf, Koivisto:2008xf,Himmetoglu:2008zp,ArmendarizPicon:2009ai}), though the spontaneous breaking of Lorentz invariance has not been the primary focus of their investigations. From this perspective, broken Lorentz invariance offers a new framework to study modifications of gravity,  and may cast some light onto theories that have been already proposed. 

The plan of the article is as follows. In Section II we generalize the coset construction of Callan, Coleman, Wess and Zumino to theories in which the group of global Lorentz transformations is spontaneously broken. In Section III we briefly review the role of the Lorentz group as an internal local symmetry group in generally covariant theories, and study the breaking of Lorentz invariance in this framework. Section IV is devoted to an  illustration of our formalism in  theories in which the rotation group remains unbroken.  We summarize our results in Section \ref{sec:Conclusions}.

\section{Broken Lorentz Invariance}
\label{sec:Broken Lorentz Invariance}

In this section  we explore how to construct theories in which the global symmetry of the action under a given Lorentz subgroup $H$ is manifest (linearly realized), but the global  symmetry under the ``broken part" of the Lorentz group $L^\uparrow_+$ is hidden (non-linearly realized). After a brief review of the Lorentz group, we first consider how to parametrize the broken part of the Lorentz group, that is, the coset $L^\uparrow_+/H$. The corresponding parameters are the Goldstone bosons of the theory. We define the action of the full Lorentz group on this set of Goldstone bosons in such a way that they transform linearly under $H$, but non-linearly under $L^\uparrow_+/H$. Initially, the transformation that we consider is internal, that is, does not affect the spacetime coordinates of the Goldstone bosons. This is the way the Lorentz group acts in generally covariant theories, which we discuss in Section \ref{sec:Coupling to Gravity}, but it is not the way it acts in theories in Minkowski spacetime, in  which the Lorentz group is a spacetime symmetry. Hence, we subsequently extend our realization of the Lorentz group to a set of spacetime transformations.

In order to write down  Lorentz-invariant theories in which the symmetry under $H$ is manifest, we need to come up with appropriate ``covariant" derivatives that transform like the Goldstone bosons themselves. As we shall see, once these covariant derivatives have been identified, the construction of actions invariant under the full Lorentz group becomes straight-forward, and simply reduces to the construction of theories in which invariance under the linearly realized $H$ is explicit. 

\subsection{The Lorentz Group}
The Lorentz group $L$ is the set of  transformations $\Lambda^a{}_b$ that leave the Minkowski metric invariant, $\eta_{ab} \Lambda^a{}_c \Lambda^b{}_d=\eta_{cd}$. Its component connected to the identity, the proper orthochronous Lorentz group $L^\uparrow_+$, is generated by rotations $\vec{J}$ and boosts $\vec{K}$, with commutation relations
\begin{subequations}\label{eq:Lorentz Lie}
\begin{align}
	[J^i,J^j]&=i\epsilon^{ij}{}_k J^k, \\
	[J^i,K^j]&=i\epsilon^{ij}{}_k K^k,\\
	[K^i,K^j]&=-i\epsilon^{ij}{}_kJ^k.
\end{align}
\end{subequations}

Any Lorentz transformation can be written as an orthochronous transformation times a parity transformation $P$, time reversal $T$ or a combination of the latter, $PT$.  These transformations define a discrete subgroup, $V\equiv\{1, P, T, PT\}$, and  the orthochronous group $L^\uparrow_+$ may be understood as the coset 
\begin{equation}
	L^\uparrow_+=L/V.
\end{equation} 
The orthochronous group is an invariant subgroup of the Lorentz group. The elements of $V$ define a map whose square is the identity,   which also preserves the commutation relations of the Lie-algebra  of the proper orthochronous group $L^\uparrow_+$,
\begin{subequations}
\begin{eqnarray}
	&P&: J^i\mapsto P J^i P^{-1}=J^i,\quad K^i\mapsto P K^i P^{-1}=-K^i \\
	&T&: J^i\mapsto T J^i T^{-1}=J^i,\quad K^i\mapsto T K^i T^{-1}=-K^i.
\end{eqnarray}
\end{subequations}
For most of this article we are concerned with the spontaneous breaking of the proper orthochronous Lorentz group $L^\uparrow_+$. 

\subsection{Coset Construction}
\label{sec:coset construction}

Suppose now  that the proper orthochronous Lorentz group $L^\uparrow_+$ (``Lorentz group" for short) is spontaneously broken down to a subgroup $H\subset L_+^\uparrow$.  In the simplest models of this kind,  the breaking occurs because the potential energy of  a vector field  has a minimum at a non-zero value of the field, in analogy with spontaneous symmetry breaking in scalar field theories with Mexican-hat potentials. Perhaps more interesting are cases in which Lorentz invariance  is broken ``dynamically," that is,  when  a strong interaction causes fermion bilinears to condense into spacetime vectors \cite{Bjorken:1963vg,Kraus:2002sa,Sannino:2002wp}.  This is analogous to the way in which chiral invariance is broken in QCD.  The formalism we develop here however does not depend on the actual mechanism that triggers the symmetry breaking, and only relies on the unbroken group $H$.

Let $\mathcal{H}$ be the Lie algebra of $H$, which we assume to be semisimple. Although the Lorentz group is not compact, it is simple, so the Killing form $(\cdot,\cdot)$ is non-degenerate and may be regarded as a scalar product on $\mathcal{H}$.  We may then uniquely decompose the Lie algebra of $L^\uparrow_+$ into the algebra of $H$ and its orthogonal complement, which we denote by $\mathcal{C}$,
\begin{equation}\label{eq:split}
	L_+^\uparrow=\mathcal{H}\oplus \mathcal{C}.
\end{equation}
Hence, by definition, for any $t\in \mathcal{H}$ and any $x\in \mathcal{C}$, $(t,x)=0$. In the following we assume that the set of unbroken generators $t^i$ is a basis of $\mathcal{H}$, and that the set of broken generators $x^m$ forms a basis of $\mathcal{C}$.   In any representation, $l^k$ collectively denotes the generators of the Lorentz group, $k=1,\ldots,6$. 

For any $t\in\mathcal{H}$, the map $f_t:x\in\mathcal{C}\mapsto [t,x]$ is linear. Moreover, for any $t'\in \mathcal{H}$ we have
\begin{equation}
	(t', [t,x] )=([t',t],x)=0,
\end{equation}
where we have used the properties of the Killing form and  that $[t,t'] \in \mathcal{H}$. Therefore, $f_t$ maps $\mathcal{C}$ into itself.\footnote{It is at this point where the assumption of a semisimple group becomes necessary. As an illustration of this point, consider the case where the unbroken group is spanned by the single generator $t \equiv K^1+J^2$. Then, the commutation relations (\ref{eq:Lorentz Lie}) imply $[t, K^3] = i t$, which is not in $\mathcal{C}$.}   In fact, the commutator defines a homomorphism of $\mathcal{H}$ into the linear maps of $\mathcal{C}$. Hence, the matrices $C(t)$ with elements defined by 
\begin{equation}\label{eq:Cartan}
	[t, x^m]=i C(t)_n{}^m \,  x^n
\end{equation}
provide a representation of $\mathcal{H}$. In particular, equation (\ref{eq:Cartan}) implies that, for any element of the unbroken group $h\in H$ and for any $x \in \mathcal{C}$,
\begin{equation}\label{eq:invariance B}
	h\,x \,h^{-1}\in \mathcal{C}.
\end{equation}

Following the standard coset construction of Callan, Coleman, Wess and Zumino \cite{Coleman:1969sm,Callan:1969sn} (see \cite{Bando:1987br, Weinberg:1996kr} for brief reviews), we can write down realizations of the Lorentz group, in which any given set of fields transform in a linear representation of the unbroken group $H$. For that purpose, let us first introduce a convenient parametrization of the coset space $L^\uparrow_+/H$. Any element $\gamma \in L^\uparrow_+/H$ can be expressed as
\begin{equation} \label{gamma}
	\gamma(\pi)=\exp(i \pi_m \, x^m),
\end{equation}
where a sum over indices in opposite locations is always implied. The fields $\pi_m=\pi_m(x)$ correspond to the Goldstone bosons of the theory. If  there are $M$ broken generators of the Lorentz group, there are $M$ Nambu-Goldstone bosons $\pi_m$.\footnote{See \cite{Low:2001bw} for exceptions to this argument in the case of spontaneous breaking of translations.} 

We may now introduce a realization of the group $L_+^\uparrow$ on this set of Goldstone bosons. By definition, any $g \in L_+^\uparrow$ can be uniquely decomposed into the product of an element of the unbroken group $h \in H$ and an element of the coset space $\gamma \in  L_+^\uparrow/H$, such that $g = \gamma \,  h$. Therefore, the product $g \, \gamma(\pi) \in L_+^\uparrow$ also has a unique decomposition
\begin{equation}\label{eq:transformation}
	 g\, \gamma(\pi(x))=\gamma(\pi'(x)) \, h(\pi(x),g), \qquad  \text{with} \qquad \gamma (\pi') \in  L_+^\uparrow/H \ , \quad h(\pi,g)\in H.
\end{equation}

Equation (\ref{eq:transformation}) defines a non-linear realization of the Lorentz group by mapping $\pi$ into $\pi'$ for any given $g \in L_+^\uparrow$. Notice however that this representation becomes linear when $g$ belongs to $H$. 
In fact, because of equation (\ref{eq:invariance B}) we must have that $\bar{h}\,\gamma(\pi) \,\bar{h}^{-1}=\gamma(\pi')$ for every $\bar{h} \in H$, and a comparison with equation (\ref{eq:transformation}) implies
\begin{equation}\label{eq:h}
	h(\pi,\bar{h}) = \bar{h}.
\end{equation}
In particular, use of equations (\ref{eq:Cartan}), (\ref{gamma}) and (\ref{eq:h}) shows that in this case the Goldstone bosons transform in  a linear representation of the unbroken group $H$,
\begin{equation}\label{eq:R}
	h\in H: \pi_m\mapsto \pi'_{\, m}=R(h)_m{}^n \, \pi_n, \quad \text{with}\quad  
	R\left(\exp i t \right) \equiv \exp\left[ i C(t) \right].
\end{equation}
Therefore, the Goldstone bosons have the same ``quantum numbers" as the broken generators $x^m$. For  a \emph{compact}, connected, semi-simple Lie group $G$ broken down to $H$, the uniqueness of the transformation law (\ref{eq:transformation}), up to field-redefinitions, was proved in \cite{Coleman:1969sm}. 

\subsection{Covariant Derivatives}
\label{sec:Covariant Derivatives}

Thus far, the realization of the Lorentz group that we have defined in equation (\ref{eq:transformation}) treats the Lorentz group as an internal symmetry; the spacetime arguments on both sides of the equation coincide.  This is going to be useful in our discussion of the Lorentz group in generally covariant theories, but it is not the way the Lorentz group acts in conventional field theories in Minkowski spacetime, in which the Lorentz group is a group of spacetime symmetries. Following \cite{Volkov:1973vd, Ogievetsky}, we define now a  non-linear realization of the Lorentz group  as a spacetime symmetry  by 
\begin{align}\label{eq:spacetime transformation}
	g: \gamma(\pi(x))\mapsto \gamma(\pi'(x')), \quad  \text{where} \quad 
	 g \, e^{iP_\mu x^\mu} \gamma(\pi(x))= e^{iP_\mu x'^\mu} \gamma(\pi'(x')) h(\pi(x),g).
\end{align}
This  implicitly defines a realization of the Lorentz group on the coordinates $x^\mu$ and the  fields $\pi(x)$.  In particular, under an arbitrary element $g\in L^\uparrow_+$ , equation  (\ref{eq:spacetime transformation}) implies
\begin{equation}
	g: x^\mu \mapsto x'^\mu =\Lambda^\mu{}_\nu(g) x^\nu,\quad 
	\gamma(\pi(x))\mapsto \gamma(\pi'(x'))=\gamma(\pi'(x)),
\end{equation}
with  $g P_\mu g^{-1}=\Lambda^\nu{}_\mu{}(g) P_\nu$ and $\gamma(\pi'(x))$ defined in equation (\ref{eq:transformation}).

Because we are interested in theories in which the Lorentz group is a set of global symmetries, any action constructed from the Goldstone bosons $\pi$ can only depend on their derivatives.  In order to introduce appropriate covariant derivatives, in analogy with the conventional prescription \cite{Callan:1969sn}, we expand  an appropriately modified  \cite{Volkov:1973vd, Ogievetsky} Maurer-Cartan form  in the basis of the Lie algebra,
\begin{equation}\label{eq:Maurer Cartan}
	\boldsymbol{\Omega}_\mu\equiv \frac{1}{i} \gamma^{-1} e^{-iP\cdot x}\partial_\mu ( e^{i P\cdot x} \gamma)
	\equiv
	e_\mu{}^a P_a+
	D_{\mu m} \, x^m+E_{\mu i} \, t^i
	\equiv  \boldsymbol{e}_\mu{}+
	\boldsymbol{D}_\mu+\boldsymbol{E}_\mu,
\end{equation}
which immediately implies that
\begin{equation}\label{eq:Minkowski vierbein}
	e_\mu{}^a=\Lambda_\mu{}^a(\gamma).
\end{equation}
The field $e_\mu{}^a$ is the analogue of the vierbein that we shall introduce in Section \ref{sec:Coupling to Gravity}. Both transform similarly under the  Lorentz group, and this leads to formally  identical expressions in both cases. But the reader should nevertheless realize that the ``vierbein" (\ref{eq:Minkowski vierbein}) and the vierbein of Section \ref{sec:Coupling to Gravity} are actually different objects.  

The transformation properties of $\boldsymbol{e}, \boldsymbol{D}$ and $\boldsymbol{E}$ follow from the definition (\ref{eq:spacetime transformation}). Under  an arbitrary $g\in L^\uparrow_+$, they transform according to
\begin{subequations}\label{eq:Minkowski e D E transf}
\begin{align}
	 g&:&\boldsymbol{e}_\mu(x)&\mapsto   \boldsymbol{e}'_\mu(x')=\Lambda_\mu{}^\nu(g)\, h(\pi,g) \boldsymbol{e}_\nu(x) h^{-1}(\pi,g), \\
	&{}& \boldsymbol{D}_\mu(x)&\mapsto   \boldsymbol{D}'_\mu(x')= \Lambda_\mu{}^\nu(g)\,  h(\pi,g)\boldsymbol{D}_\nu(x) h^{-1}(\pi,g), \\
	&{}&  \boldsymbol{E}_\mu(x)&\mapsto \boldsymbol{E}'_\mu(x')= \Lambda_\mu{}^\nu(g)  \left[h(\pi,g)\boldsymbol{E}_\nu (x) h^{-1}(\pi,g)	- i h(\pi,g)\partial_\nu h^{-1}(\pi,g)\right],
\end{align}
\end{subequations}
where $h(\pi,g)$ is defined in equation (\ref{eq:transformation}). Therefore, none of these quantities really transforms covariantly, since the spacetime index $\mu$  and the components of the different fields transform under different group elements.  To define fully covariant quantities, let us introduce  the inverse of the quantity defined in equation (\ref{eq:Minkowski vierbein}),
\begin{equation}
	e^\mu{}_a=\Lambda_a{}^\mu(\gamma^{-1}). 
\end{equation} 
This is indeed the (transposed) inverse of $e_\mu{}^a$ because it follows from equation (\ref{eq:Minkowski vierbein}) that
$e^\mu{}_a e_\mu{}^b =\delta_a{}^b$.   Then, the quantities
\begin{equation}\label{eq:Minkowski Da}
	\boldsymbol{\mathcal{D}}_a\equiv e^\mu{}_a \boldsymbol{D}_\mu,
	\quad
	 \boldsymbol{\mathcal{E}}_a\equiv e^\mu{}_a \boldsymbol{E}_\mu,
\end{equation}
do transform covariantly under the Lorentz group,
\begin{subequations}\label{eq:E D transf}
\begin{align}
	\boldsymbol{\mathcal{D}}_a(x)&\mapsto \boldsymbol{\mathcal{D}}'_a(x')=\Lambda(h(\pi,g))_a{}^b\, 
	h(\pi,g)\boldsymbol{\mathcal{D}}_b(x) h^{-1}(\pi,g), \\
	\boldsymbol{\mathcal{E}}_a(x)&\mapsto \boldsymbol{\mathcal{E}}'_a(x')=\Lambda(h(\pi,g))_a{}^b 
	\left[ h(\pi,g)\boldsymbol{\mathcal{E}}_b(x) h^{-1}(\pi,g)
	- i  h(\pi,g) \partial_bh^{-1}(\pi,g)\right],
\end{align}
\end{subequations}
where $\partial_a\equiv e^\mu{}_a \partial_\mu$. We identify $\boldsymbol{\mathcal{D}}_a$ with the covariant derivative of the Goldstone bosons $\pi_m$, and $\boldsymbol{\mathcal{E}}_{a}$ with a ``gauge field" that  will enter the couplings between the Goldstone bosons and other matter fields.  The transformation rules (\ref{eq:E D transf}) are again non-linear in general, but, because of equation (\ref{eq:h}),  they reduce to a linear transformation if $g\in H$. Note that under $g\in L^\uparrow_+$, the components of the covariant derivative $\boldsymbol{\mathcal{D}}_a$ transform as
\begin{equation}\label{eq:tensor}
	g: \mathcal{D}_{a m}(x)\mapsto \mathcal{D}'_{a m} (x')=\Lambda_a{}^b(h(\pi,g)) R_m{}^n \mathcal{D}_{b n}(x),
\end{equation}
where the matrix $R$ is the one we introduced in equation (\ref{eq:R}). 

For specific calculations, it is often required to have concrete expressions for the covariant derivatives. It follows from the definitions (\ref{gamma}) and (\ref{eq:Maurer Cartan}) that
\begin{equation}\label{eq:concrete covD}
	\mathcal{D}_{a m}=\partial_a \pi_m - i \pi_n (x^n_{(4)})_a{}^b \partial_b \, \pi_m+\frac{1}{2}\pi_n \partial_a \pi_p \, C^{np}{}_m + \mathcal{O}(\pi^3),
\end{equation}
where $x^n_{(4)}$ is the fundamental (form) representation of the Lorentz generator $x^n$, and the $C^{np}{}_m$ are the structure constants of the Lie algebra $\mathcal{H}$ in our basis of generators. 

\subsubsection*{Parity and Time Reversal}

In certain cases, we can also define the transformation properties of the Goldstone bosons under parity and time reversal, or, in general, under an appropriate subgroup of  ${V\equiv\{1,P, T, PT\}}$. Let $V_H$ denote the ``stabilizer" of $H$, that is, the set of all elements $v\in V$  that leave $H$ invariant,  $v \, h \, v^{-1} \in H$ for all $h\in H$.  This is a subgroup of $V,$ which  may contain just the identity,  either $P$ or $T$, or $V$ itself. Because $\mathcal{H}$ is invariant under $V_H$,  the latter  defines an homomorphism on $\mathcal{C}$ by conjugation,
\begin{equation}
	v\,  x^m v^{-1} =V_n{}^m \, x^n.
\end{equation}

The two sets  $L^\uparrow_+ V_H$ and $H V_H$ are two  subgroups of $L$,  and, by definition,  $H V_H$ is a subgroup of $L^\uparrow_+ V_H$. Thus, just as in Section \ref{sec:coset construction} , we may define a realization of $V_H$  (which is now contained in  $L^\uparrow_+ V_H$) on the coset 
\begin{equation}
	\frac{L^\uparrow_+ V_H}{H V_H}=\frac{L^\uparrow_+}{H}.
\end{equation}
In particular, for $g\in L^\uparrow_+ V_H$ and $\gamma(\pi)\in L^\uparrow_+/H$ we set 
\begin{equation}
		g\gamma(\pi)=\gamma(\pi')h(\gamma,g) v(\gamma,g), \quad
		\text{with}
		\quad
		h(\gamma,g)\in H \,\, \text{and}\,\,	v(\gamma,g)\in V_H. 
\end{equation}
If $g\in L^\uparrow_+$, this definition reduces to that of equation (\ref{eq:transformation}). For $v\in V_H$ it leads to
\begin{equation}\label{eq:gamma transf V}
	v:\gamma(\pi)\mapsto \gamma(\pi')=v\, \gamma(\pi) \, v^{-1},
\end{equation}
which can be extended to include the arguments of the Goldstone boson fields as before,
\begin{equation}
	v: \gamma(\pi(x))\mapsto \gamma(\pi'(x')), \quad \text{where} \quad 
	v \, e^{i P\cdot x} \gamma(\pi(x))\,  v^{-1}= e^{i P\cdot x'} \gamma(\pi'(x')).
\end{equation}
Under these group elements the Goldstone bosons change according to 
\begin{equation}\label{eq:X law pi}
	v: \pi_m\mapsto \pi'_m(x')=V_m{}^n \, \pi_n(x), 
\end{equation} 
and, from equation (\ref{eq:tensor}), their covariant derivatives according to
\begin{equation}
	v:  D_{a m}(x)\mapsto D'_{a m}(x')=V_a{}^b V_m{}^n D_{b n}(x),
\end{equation}
where  $v P_a v^{-1}= V_a{}^b P_b$. 

\subsection{Invariant Action}

If we are interested in the low-energy limit of theories in which Lorentz-invariance is broken, we can restrict our attention to their massless excitations. This is a restatement of the  Appelquist-Carazzone theorem \cite{Appelquist:1974tg},  though the latter  has been actually proven  only  for renormalizable Lorentz-invariant theories in flat spacetime. Typically, massless fields are those protected by a symmetry, and always include the Goldstone bosons, since invariance under the broken symmetry prevents them from entering the action undifferentiated. Therefore, the low-energy effective action of any theory in which Lorentz invariance is broken must contain the covariant derivatives of the Goldstone bosons. To leading order in the low-energy expansion, we can restrict our attention to the minimum number of spacetime derivatives, namely, two.

The tensor product representation in equation (\ref{eq:tensor}) under which the covariant derivatives transform is in general reducible. Let ${\Lambda\otimes R=\oplus_{i} R^{(i)}}$ be its Clebsch-Gordan series, and let $\mathcal{D}^{(i)}$ be the linear combination of covariant derivatives that furnishes the $i$-th irreducible representation. Some of these representations may be singlets, and we shall label them by $s$.  Because the unbroken group is not necessarily compact,  the non-trivial irreducible representations are generally not unitary. In any case, if $G^{(i)}$ is invariant under the $i$-th representation of the unbroken group $H$, i.e. $R^{(i)}{}^T G^{(i)} R^{(i)}=G^{(i)}$, then the Lagrangian
\begin{equation}\label{eq:Goldstone invariants}
	\mathcal{L}=\sum_s F_s \mathcal{D}^{(s)}+\sum_i  F_i \, \mathcal{D}^{(i)}{}^T G^{(i)} \mathcal{D}^{(i)} 
\end{equation}
transforms as a scalar under the Lorentz-group $L_+^\uparrow$. Here, $F_s$ and $F_i$ are free parameters in the effective, which remain undetermined by the symmetries of the theory. In order to construct a Lorentz-invariant action, we just need a volume element that transforms appropriately under our realization of the Lorentz group. This is in general given by \cite{Ogievetsky}
\begin{equation}\label{eq:dV}
	d^4V\equiv d^4x \det e_\mu{}^a,
\end{equation}
which, because of equation (\ref{eq:Minkowski vierbein}), results in $d^4 V=d^4x$. (Inside the determinant, the vierbein should be regarded as a $4\times 4$ matrix with rows labeled by $\mu$ and columns labeled by $a$.) The functional
\begin{equation}
	S=\int d^4V \mathcal{L}
\end{equation}
is then invariant under the action of the Lorentz group defined by equation (\ref{eq:spacetime transformation}).

\subsection{Couplings to Matter}

The formalism can be also extended to capture the effects of Lorentz breaking on the matter sector. As mentioned above, at low-energies we can restrict our attention to massless (or light) fields, which are typically those that are prevented from developing a mass by a symmetry like chiral or gauge invariance. We consider couplings to the graviton in Section  \ref{sec:Coupling to Gravity}.

Let $\psi$ be any  matter field that transforms under any (possibly reducible) representation $\mathscr{R}(h)$ of the unbroken  group $H$, with generators $t_i$. Let us now \emph{define} the transformation law under the full Lorentz group to be \cite{Coleman:1969sm}
\begin{equation}\label{eq:matter tilde transformation}
	g:\psi(x)\mapsto\psi'(x')=\mathscr{R}(h(\pi,g)) \,\psi(x),
\end{equation}
where $x'$ and $h(\pi,g)$ is given in equation (\ref{eq:spacetime transformation}). We can also construct  covariant derivatives  under the Lorentz group by setting,
\begin{equation}\label{eq:matter tilde covD}
	\mathscr{D}_a \psi \equiv  e^\mu{}_a \left[ \partial_\mu \psi
	+i \boldsymbol{E}_{\mu}\psi\right]= \partial_a \psi+i \boldsymbol{\mathcal{E}}_a\psi,
\end{equation}
where $\boldsymbol{E}_\mu$ is defined in equation (\ref{eq:Maurer Cartan}). The covariant derivative transforms just as the field itself, under a representation of the same group element,
\begin{equation}
	g: \mathscr{D}_a \psi(x) \mapsto \mathscr{D}'_a \psi'(x')=\Lambda(h(\pi,g))_a{}^b \mathscr{R}(h(\pi,g)) \,
	\mathscr{D}_b\psi (x).
\end{equation}
Therefore, any  Lagrangian built out of $d^4V$, $\psi$, $\mathscr{D}_a \psi$ and $\mathcal{D}_{a m}$ that is invariant under the unbroken group $H$ is then invariant under the full Lorentz group.

With these ingredients we could develop a formulation of the standard model in which the Lorentz group is spontaneously broken to any subgroup. If the unbroken group is trivial, $H=1$, this construction would be analogous to the standard model extension considered by Colladay and Kostelecky \cite{Colladay:1998fq}. Our article mainly focuses on the general formalism of broken Lorentz invariance,  so we shall not carry out this program here. For the purpose of illustration however, and in order to establish the connection to previous  work on the subject, let us consider a formulation of QED (quantum electro-dynamics) in which the Lorentz group is completely broken. For simplicity we consider a theory with a single ``spinor" $\psi_\alpha$ of charge $q$ coupled to a ``photon" $A_a$. We use quotation marks because, according to (\ref{eq:matter tilde transformation}), we assume that under the (completely) broken Lorentz group both fields are invariant. On the other hand, we require that the theory be invariant under gauge transformations, that is, we demand invariance under
\begin{equation}
	\psi_\alpha\to e^{i q \chi} \psi_\alpha,
	\quad 
	A_a\to A_a+\partial_a\chi,
\end{equation}
where $\chi$ is an arbitrary spacetime scalar. If the Lorentz group is broken down to $H=1$, there are six Goldstone bosons in the theory, and $\gamma$ becomes $\gamma\equiv \exp(i\pi^k l_k)$, which, under the Lorentz group transforms as
${	g: \gamma\mapsto \gamma'(x')=g \, \gamma(x).}$
Following (\ref{eq:matter tilde covD}) we introduce now the covariant derivatives
\begin{equation}
	\mathscr{D}_a A_b\equiv \Lambda(\gamma^{-1})_a{}^\mu \partial_\mu A_b,
	\qquad
	\mathscr{D}_a\psi_\alpha\equiv \Lambda(\gamma^{-1})_a{}^\mu 
	\partial_\mu\psi_\alpha,
\end{equation}
which by construction are Lorentz-invariant (if the Lorentz group is completely broken, $\boldsymbol{E}_{\mu}\equiv 0$ by definition.) Gauge invariance then dictates that the derivatives of the fields must enter in the gauge invariant or covariant forms
\begin{equation}
	F_{ab}\equiv
	\mathscr{D}_a A_b-\mathscr{D}_a A_b, 
	\quad
	\nabla_a \psi_\alpha\equiv 
	(\mathscr{D}_a-iq A_a)\psi_\alpha.
\end{equation}
Any gauge invariant combination of these elements, such as 
\begin{equation}\label{eq:QEDext}
	\mathcal{L}_{QED}=M^{abcd}F_{ab}F_{cd}+
	N^{\alpha\beta}\psi^\dag_\alpha \nabla_a\psi_\beta
	+P^{\alpha\beta} \psi^\dag_\alpha \psi_\beta,
\end{equation}
is also Lorentz invariant (for simplicity, we have not written down all the terms compatible with the two symmetries). In equation (\ref{eq:QEDext}),  the dimensionless matrices $M$, $N$ and $P$ are \emph{arbitrary}, up to the restrictions imposed by permutation symmetry and hermiticity. The Lagrangian (\ref{eq:QEDext}) is thus the analogue of the extension of QED described in \cite{Colladay:1998fq}. From a phenomenological perspective, its coefficients can be regarded as quantities to be determined or constrained by experiment, as in the standard model extension of \cite{Colladay:1998fq}.  But of course, as opposed to the latter,  the Lagrangian (\ref{eq:QEDext}) contains couplings to the Goldstone bosons, and  should be supplemented with the Goldstone boson Lagrangian, which for a trivial $H$ is
\begin{equation}
	\mathcal{L}_\pi= G^{am} D_{am}+F^{mnab}\, \mathcal{D}_{am} \mathcal{D}_{b n},
\end{equation}
where $\mathcal{D}_{a m}$ is given in equation (\ref{eq:Maurer Cartan}), and $m, n=1,\ldots, 6$. As we shall see in the next section, in a gravitational theory these covariant derivatives should be included in the Lagrangian too, but in that case they reduce to  appropriate components of the spin connection.  Note that in our conventions the Goldstone bosons are dimensionless. Thus the coefficients in $G$ have mass dimension three, and those in $F$ mass dimension two. In theories in which an internal symmetry is spontaneously broken, Lorentz invariance and invariance under the unbroken group severely restrict the possible different mass scales appearing in the Lagrangian. In our case however, the values of $G$ and $F$ are (up to symmetry under permutations) completely arbitrary. In particular, the unbroken symmetries do not imply that there is a single energy scale at which the Lorentz group is broken.

The obvious problem with this approach is that the Lorentz group seems to be an unbroken symmetry in the matter sector.   A generic Lagrangian like (\ref{eq:QEDext}), constructed out of the standard model fields $\psi$, their covariant derivatives $\mathscr{D}_a \psi$ and the covariant derivatives of the Goldstones $\mathcal{D}_{a m}$ would  clearly violate Lorentz invariance, in flagrant conflict with experimental constraints \cite{Kostelecky:2008ts}.  Thus, we are forced to assume that these ``Lorentz-violating" terms are sufficiently suppressed, which in our context requires specific relations between the coefficients in the effective Lagrangian.

To illustrate this point, let us briefly discuss how to construct scalars  under linearly realized Lorentz transformations out of the ingredients at our disposal, namely, $\psi, \mathscr{D}_a \psi$ and $\mathcal{D}_{a m}$.   Imagine that the matter fields $\tilde{\psi}$ actually fit in a representation of the Lorentz group $\mathscr{R}(g)$. It is then more convenient to postulate that under the \emph{full} Lorentz group, these fields transform as
\begin{equation}\label{eq:linear psi}
	g: \tilde{\psi}\mapsto \tilde{\psi}'(x')=\mathscr{R}(g)\tilde{\psi}(x).
\end{equation}
Then, any Lagrangian that is invariant (a scalar) under \emph{global} Lorentz transformations,
\begin{equation}\label{eq:L inv}
	\mathcal{L}_\mathrm{inv}[\tilde{\psi},\partial_\mu\tilde{\psi}]=
	\mathcal{L}_\mathrm{inv}[\mathscr{R}(g) \tilde{\psi}, \mathscr{R}(g)\Lambda(g)_\mu{}^\nu\partial_\nu\tilde{\psi}], \quad g\in L^\uparrow_+,
\end{equation}
is clearly invariant under the unbroken subgroup $H$ of global transformations, and can thus be part of the effective Lagrangian in the spontaneously broken phase. Note that these Lorentz invariant terms would not contain any couplings to the Goldstone bosons. But given the transformation law (\ref{eq:linear psi}) we can also construct a new quantity that transforms under the non-linear realization of the Lorentz group (\ref{eq:matter tilde transformation}),
\begin{equation}\label{eq:psi tilde psi}
	\psi\equiv \mathscr{R}(\gamma^{-1})\tilde{\psi},
\end{equation}
and whose covariant derivative can again be defined by equation (\ref{eq:matter tilde covD}). In this case, however, the field $\psi$ is to be understood simply as a shorthand for the right hand of equation (\ref{eq:psi tilde psi}), which contains the Goldstone bosons $\gamma(\pi)$. Given any Lagrangian $\mathcal{L}_\mathrm{break}$ that is invariant under the linearly realized unbroken group $H$, but not invariant under linear representations of the full Lorentz group $L^\uparrow_+$,
\begin{equation}
	\mathcal{L}_\mathrm{break}[\psi, \partial_\mu\psi]=
	\mathcal{L}_\mathrm{break}[\mathscr{R}(h) \psi, \mathscr{R}(h)\Lambda(h)_\mu{}^\nu \partial_\nu\psi],
	\quad h\in H,
\end{equation}
 we can then construct further invariants under Lorentz transformations,
\begin{equation}\label{eq:L break}
	\mathcal{L}_\mathrm{break}[\mathscr{R}(\gamma^{-1})\tilde{\psi}, \mathscr{D}_\mu (\mathscr{R}(\gamma^{-1})\tilde{\psi})].
\end{equation}
Here, the appearance of the Goldstone bosons in those terms that violate the full Lorentz symmetry is  manifest.

It seems now that the Lagrangians (\ref{eq:L inv}) and (\ref{eq:L break}) do not fit into the general prescription to construct invariant Lagrangians that  we described at the beginning of this subsection, but this is just an appearance. Suppose we perform a field redefinition $\mathscr{R}(\gamma^{-1})\tilde{\psi}\to \psi$, and assume that the new field $\psi$ transforms as in equation (\ref{eq:matter tilde transformation}).  This field redefinition turns the Lagrangian in equation  (\ref{eq:L break}) into  $\mathcal{L}_\mathrm{break}[\psi, \mathscr{D}_\mu \psi]$,
and takes the Lagrangian (\ref{eq:L inv}) into
\begin{equation}\label{eq:L tilde inv L}
	\mathcal{L}_\mathrm{inv}[\psi, \mathscr{D}_{\mu}\psi+i D_{\mu m} x^m \psi].
\end{equation}
Both Lagrangians are invariant under the linearly realized symmetry group $H$ (and the non-linearly realized Lorentz group $L^\uparrow_+)$, and both are solely constructed in terms of  $\psi, \mathscr{D}_\mu\psi$ and $\mathcal{D}_{a m}$.

Of course, a general Lagrangian invariant under $H$ will have the form of equation (\ref{eq:L tilde inv L}) only for very particular choices of the coefficients that remain undetermined under the unbroken symmetry. From the point of view of the effective theory, this particular choice cannot be explained, though it is certainly compatible with the symmetries we are enforcing. To address it we would have to rely on specific models. Say, if Lorentz symmetry is broken in a hidden sector which is completely decoupled from the standard model, the breaking in the hidden sector should not have any impact on the visible sector. But of course, the two sectors must couple at least gravitationally. Then,  if the scale of Lorentz-symmetry breaking is sufficiently small compared to the Planck mass, we expect a double   suppression of Lorentz-violating terms in the matter sector: from the weakness of gravity, and from the smallness of the symmetry breaking scale. We defer the discussion of gravitation to the next section.  Radiative corrections to Lorentz-violating couplings in the matter sector  of Einstein-aether models \cite{Jacobson:2000xp} have been calculated in \cite{Withers:2009qg}.

\subsection{Broken Rotations}
\label{sec:Broken rotations}

As an example of the formalism discussed so far, we shall briefly study a pattern of symmetry breaking  in which the unbroken group $H$ is non-compact. This is an interesting case since, for internal non-compact symmetry groups,  the theory  contains  ghosts in the spectrum of Goldstone bosons \cite{Bando:1987br,Kraus:2002sa}. We show that, instead, it is certainly possible to have a well-behaved spectrum in a theory in which the Lorentz group is broken down to a non-compact subgroup. We consider the widely-studied case of unbroken rotations, $H=SO(3)$, in Section \ref{sec:Unbroken rotations}.

Suppose  that the Lorentz group $L_+^\uparrow$ is broken down to the group of transformations that leave the vector field $A^\mu=(0,0,0,F)$ invariant. This breaking pattern was studied in references \cite{Bjorken:1963vg,Kraus:2002sa}, in which the  ``photon" of electromagnetism is identified with the Goldstone bosons associated with the breaking. The Lie algebra of the unbroken group $H$ is then
\begin{equation}
	\mathcal{H}=\text{Span}\{K^1,K^2,J^3\},
\end{equation}
which is simple, and isomorphic to the Lie algebra of the group of Lorentz transformations in three-dimensional spacetime $so(1,2)$. Its orthogonal complement is spanned by
\begin{equation}
	\mathcal{C}=\text{Span}\{J^1,J^2,K^3\}.
\end{equation}
Because $\text{dim}(\mathcal{C})=3$, there are three Goldstone bosons in the theory. It follows from the commutation relations (\ref{eq:Lorentz Lie}) and equations (\ref{eq:Cartan}) and (\ref{eq:R}) that $\pi_m\equiv (\pi_3,\pi_1,\pi_2)$ transforms like a Lorentz three-vector. It is thus convenient to let $m$ run from $0$ to $2$ and identify $\pi_0\equiv \pi_3$.

The covariant derivative $\mathcal{D}_{a m}$ transforms in a  reducible representation of ${H=SO(1,2)}$. The covariant derivative 
\begin{equation} \label{covderSO(2,1)}
	\mathcal{D}_m\equiv \mathcal{D}_{3 m}
\end{equation}
is an $SO(1,2)$ three-vector. The remaining irreducible spaces are spanned by the scalar $\vphi$, the vector $a_{mn}$ and the tensor $s_{mn}$ defined by
\begin{equation} \label{DmnSO(2,1)}
	\vphi\equiv \mathcal{D}_m{}^m,\, \,  
	a_{mn}\equiv\frac{1}{2}(\mathcal{D}_{mn}-\mathcal{D}_{nm}), \, \,
	s_{mn}\equiv  \frac{1}{2}(\mathcal{D}_{mn}+\mathcal{D}_{nm})-\frac{1}{3}\vphi\, \eta_{mn},
\end{equation}
where  indices are raised with the (inverse) of the Minkowski metric in three dimensions, $\eta^{mn}=\mathrm{diag}(-1,1,1)$ and $m=0,1,2$. Scalar invariants are constructed then by appropriate contraction of indices,
\begin{equation}\label{eq:SO(1,2) invariant}
	\mathcal{L}_\pi= G_\vphi \vphi+ F_\vphi\, \vphi^2+F_D  \mathcal{D}_m \mathcal{D}^m+F_a\,  a_{mn} a^{mn}
	+F_{\epsilon}\,  \epsilon_{mnp}\, a^{mn} \mathcal{D}^p
	+F_s \, s_{mn} s^{mn}.
\end{equation}

For simplicity, let us now consider the case where $G_\vphi = 0$. Because to lowest order in the Goldstone bosons $\mathcal{D}_{mn}=\partial_m \pi_n+\cdots$, inspection of (\ref{eq:SO(1,2) invariant}) reveals the lower-dimension analogue of a generalized vector field theory in which the vector field consists of the Goldstone bosons $\pi_m$. This analogy can be further strengthened by dimensionally reducing the four dimensional theory from four to three spacetime dimensions. Expanding the Goldstone bosons in Kaluza-Klein modes
\begin{equation}
	\pi_m(t,x,y,z)=\sum_{k_z}\pi^{(k)}_m(t,x,y) e^{i k z},
\end{equation}
and inserting into the action we obtain, to quadratic order, 
\begin{eqnarray}
	S&=&\sum_{k} S_{k}, \quad \text{where} \nonumber \\
	 S_k [\pi^{(k)}_m]&=&\int dt \, d^2x \left[\frac{F_a+F_s}{4} \, a_{mn} a^{mn}
	+\left(F_\vphi+\frac{2F_s}{3}\right) (\partial_m \pi^m)^2
	+F_D \, k^2 \, \pi_m \pi^m\right]. \label{eq:3d aether}
\end{eqnarray}
Note that we have suppressed the index $k$ of the Kaluza-Klein modes on the right hand side of equation (\ref{eq:3d aether}). The Kaluza-Klein modes $\pi^{(k=0)}$ are massless, and transform like an $SO(1,2)$ vector. They can be thought of as the Goldstone bosons associated with the breaking $L_+^\uparrow\sim SO(1,3)\to SO(1,2)$ induced by the compactification. 

The spectrum of excitations in the theory described by (\ref{eq:3d aether}), and the conditions that stability imposes on the free parameters $F_a, F_\vphi$, $F_s$ and $F_D$ can be derived by relying on the similarity of the action $S_{k}$ with the four-dimensional models analyzed in \cite{ArmendarizPicon:2009ai}. Since their stability analysis does not crucially depend on the dimensionality of spacetime, their results also apply in the case at hand.\footnote{There is just  one difference between the four-dimensional and the three-dimensional case: In four dimensions, the vector sector (under spatial rotations) contains two modes, while in three dimension the vector sector (under spatial rotations) only contains one mode.} Following the analysis in Section V of \cite{ArmendarizPicon:2009ai}  we find:
\begin{enumerate}
	\item [i)]If both $F_a +F_s$ and $F_\vphi+2F_s/3$ are different from zero, the spectrum consists of an $SO(1,2)$ vector and an $SO(1,2)$ scalar. There is always a ghost at high spatial momenta ($k_x^2+k_y^2\gg k^2$).
	\item[ii)] For $F_a+F_s=0$ the theory is stable if $F_\vphi+2F_s/3>0$ and $F_D<0$. The spectrum consists of a scalar under $SO(1,2)$. If $F_D k^2=0$, there are no dynamical fields in the spectrum.
	\item[iii)] For $F_\vphi+2F_s/3=0$ the Lagrangian is the three-dimensional version of the Proca Lagrangian. The spectrum consists of a massive $SO(1,2)$ vector, with two polarizations. The theory is stable for $F_a+F_s>0$ and $F_D<0$. If $F_D k^2=0$ the vector is massless, with only one polarization. This last cast corresponds to electrodynamics in three spacetime dimensions.
\end{enumerate}
Hence, as we anticipated there are theories in which the low-energy theory is free of ghosts. These are however non-generic, in the sense that they require the coefficients of certain terms otherwise allowed by Lorentz invariance to be zero.

\section{Coupling to Gravity}
\label{sec:Coupling to Gravity}

In the previous section we have explored spontaneous symmetry breaking of Lorentz invariance in Minkowski spacetime, in which the Lorentz group is a global symmetry. Though this approach should appropriately capture the local physical implications of the breaking in  non-gravitational phenomena, it certainly does not suffice to study  arbitrary spacetime backgrounds, or  the gravitational interactions themselves.

In order to couple gravity to the Goldstone bosons, it is convenient to exploit the formal analogy between gravity and gauge theories. For that purpose, one introduces the Lorentz group $L^\uparrow_+$ as an ``internal" group of symmetries, in addition to the symmetry under general coordinate transformations \cite{Utiyama:1956sy}. In theories with fermions (such as the standard model) this is actually mandatory, as the group of general coordinate transformations does not admit spinor representations.  In the first part of this section we review the standard formulation of general relativity as a gauge theory of the Lorentz group \cite{carroll}. In the second part we then extend this standard formulation to theories in which Lorentz invariance is broken. Readers already familiar with the vierbein formalism can skip directly to Subsection \ref{sec:Broken Lorentz Symmetry}.

\subsection{General Formalism}

In any generally covariant theory defined on a spacetime manifold in which the metric has Lorentzian signature, and regardless of  whether the Lorentz group is spontaneously broken or not, it is always possible to introduce a vierbein, an orthonormal set of forms $\hat{e}^{(a)} = e_\mu{}^a \, dx^\mu$ in the cotangent space of the spacetime manifold, 
\begin{equation}\label{eq:orthonormality}
	g^{\mu\nu}e_\mu{}^{a} \, e_\nu{}^b=\eta^{ab}.
\end{equation}
Greek indices $\mu,\nu,\ldots $ now denote cotangent space indices in a coordinate basis, while latin indices $a,b,\ldots$ label the different vectors in the orthonormal set. Thus, the order of the vierbein indices is important. The first one is always a spacetime index, and the second one is always a Lorentz index. Spacetime indices are raised and lowered with the metric of spacetime, and Lorentz indices are raised and lowered with the Minkowski metric. Under coordinate transformations, the vierbein $e_\mu{}^a$ transforms like a vector, 
\begin{equation}
	\mathrm{diff}: e_\mu{}^a(x) \mapsto e'_\mu{}^a(x')=\frac{\partial x^\nu}{\partial x'^\mu} e_\nu{}^a(x).
\end{equation}

The freedom to choose  a  vierbein whose sixteen components satisfy the orthonormality condition (\ref{eq:orthonormality}) does not  add anything to the original ten metric components  if the theory remains invariant under the six parameter group of local Lorentz transformations,
\begin{equation}\label{eq:e transf}
	g(x): e_\mu{}^a(x) \mapsto  e'_\mu{}^a(x)=\Lambda^a{}_b(g) \, e_\mu{}^b(x), \quad g(x)\in L_+^\uparrow.
\end{equation}
Note that this transformation does not affect the coordinates of the vectors, that is, the Lorentz group acts as an ``internal" symmetry. 

The derivatives of the vierbein do not transform covariantly under these local Lorentz transformations. We thus  introduce the spin connection $\boldsymbol{\omega}_\mu$, which plays the role of the gauge field of the Lorentz group.  Let $l^k$, $k=1,\ldots 6$, denote the generators of the Lorentz group $L^\uparrow_+$ (in any representation), and let us define the components of the spin connection by 
\begin{equation}
	\boldsymbol{\omega}_\mu{}\equiv \omega_\mu{}_ k \, l^k,
\end{equation}
which transforms like a one-form under general coordinate transformations,
\begin{equation}\label{eq:omega transformation}
	\mathrm{diff}: \boldsymbol{\omega}_\mu(x)\mapsto \boldsymbol{\omega}'_\mu(x')=
	\frac{\partial x^\nu}{\partial x'^\mu}\boldsymbol{\omega}_\nu(x).
\end{equation}
In complete analogy with gauge field theories, let us assume that under local Lorentz transformations the spin connection transforms  as\footnote{To recover expressions fully analogous to those found in gauge theories, the reader should replace $\boldsymbol{\omega}_{\mu}$ by $-i\boldsymbol{\omega}_{\mu}$.}
\begin{equation} \label{LLTomega}
	g(x): \boldsymbol{\omega}_\mu(x)\mapsto g \, \boldsymbol{\omega}_\mu(x) \, g^{-1}
	+ g\,  \partial_\mu g^{-1}.
\end{equation}
In that case, it is then easy to verify that the covariant derivative 
\begin{equation}\label{eq:covD e}
 	\nabla_\mu e_\nu{}^a=\partial_\mu e_\nu{}^a
	-\Gamma^\rho{}_{\nu\mu}e_\rho{}^a
	+\omega_{\mu k}\, [l_{(4)}^k]^a{}_b \, e_\rho{}^b
 \end{equation}
transforms covariantly both under coordinate and local Lorentz transformations. Here, $\Gamma^{\mu}{}_{\nu\rho}$ are the Christoffel symbols associated with the spacetime metric $g_{\mu\nu}$, and the $l_{(4)}^k$ are the Lorentz group generators  in the fundamental  representation, under which the vierbein transforms. In our convention, these matrices are purely imaginary. Similarly, given any matter field $\psi$ that transforms as a scalar under diffeomorphisms, and in a representation of the Lorentz group with generators $l_k$, we can construct its covariant derivative  by 
\begin{equation}
	\nabla_\mu \psi\equiv \partial_\mu \psi+\boldsymbol{\omega}_\mu \psi,
\end{equation}
which also transforms covariantly both under diffeomorphisms and local Lorentz transformations. 

In any generally covariant theory defined on a Riemannian spacetime manifold, the covariant derivative is compatible with the metric, that is, $\nabla_\mu g_{\nu\rho}=0$. Moreover, because the Minkowski metric is invariant under Lorentz-transformations, its  Lorentz-covariant derivative vanishes. Thus, differentiating equation (\ref{eq:orthonormality}) covariantly,  and using Leibniz rule we obtain
\begin{equation}\label{eq:covD e zero}
	\nabla_\nu e_\mu{}^a=0.
\end{equation}
Equation (\ref{eq:covD e}), in combination with equation (\ref{eq:covD e zero}) allows us to express the spin connection in terms of the vierbein,
\begin{equation}\label{eq:omega(e)}
	\omega_{\mu k} [l_{(4)}^k]^a{}_b=\frac{1}{2}\left[
	e^\nu{}^a(\partial_\mu e_\nu{}_b-\partial_\nu e_\mu{}_b)
	-e^\nu{}_b(\partial_\mu e_\nu{}^a-\partial_\nu e_\mu{}^a)
	-e^{\rho a} e^\sigma{}_ b(\partial_\rho e_{\sigma c}-\partial_\sigma e_{\rho c})e_\mu{}^c \right],
\end{equation}
and it is readily verified that this connection indeed transforms as in equation (\ref{eq:omega transformation}). 

Equation (\ref{eq:omega(e)}) is what sets gauge theories and gravity apart.  In gauge theories, the gauge fields are ``fundamental" fields on which the action functional depends. In gravity  the spin connection can be expressed in terms of the vierbein, which constitute the fundamental fields in the gravitational sector. In particular, the metric can be also expressed in terms of the vierbein,
\begin{equation}
	g_{\mu\nu}=e_{\mu a} e_\nu{}^a,
\end{equation}
where, as in Subsection \ref{sec:Covariant Derivatives},  $e^\mu{}_a$ is the (transposed) inverse of $e_\mu{}^a$, that is,
$e_\mu{}^a \, e^\nu{}_a=\delta^\mu{}_\nu$.
Because the covariant derivative of the vierbein vanishes by construction, one can use the vierbein to freely alter the transformation properties of any field under diffeomorphisms and Lorentz transformations. For instance, ${\nabla_\mu A_a\equiv e^\nu{}_a \nabla_\mu A_\nu}$, so one can use the vierbein to freely convert   diffeomorphism vectors into Lorentz vectors and vice versa.

Since the spin connection transforms like a gauge field, the curvature tensor
\begin{equation}\label{eq:Riemann}
	\boldsymbol{R}_{\mu\nu}\equiv \partial_\mu \boldsymbol{\omega}_\nu
	-\partial_\nu \boldsymbol{\omega}_\mu
	+[\boldsymbol{\omega}_\mu,\boldsymbol{\omega}_\nu]
\end{equation}
transforms like a two-form under general coordinate transformations, and in the adjoint representation under local Lorentz transformations $g(x) \in L_+^\uparrow$,
\begin{equation}
	g(x):\boldsymbol{R}_{\mu\nu}\mapsto g\, \boldsymbol{R}_{\mu\nu} \, g^{-1}.
\end{equation}
This transformation law is particularly simple in the fundamental (form) representation of the Lorentz group. In that case, for fixed $\mu$ and $\nu$ the curvature $\mathbf{R}_{\mu\nu}$  is a matrix $[R_{\mu\nu}]_a{}^b$ whose elements  transform according to
\begin{equation}
	g(x): [R_{\mu\nu}]^a{}_b \to [R'_{\mu\nu}]^a{}_b =
	\Lambda^a{}_c(g)
	\Lambda_b{}^d(g) 
	[R_{\mu\nu}]^c{}_d.
\end{equation}
Note that the curvature tensor is antisymmetric in the coordinate and Lorentz indices,
\begin{equation}\label{eq:antisymmetry}
	R_{\mu\nu a b}=-R_{\nu\mu a b}=-R_{\mu\nu b a}.
\end{equation}
Recall that spacetime indices are raised and lowered with the spacetime metric $g_{\mu\nu}$, and Lorentz indices are raised and lowered with the Minkowski metric $\eta_{ab}$.

With these ingredients it is then possible to construct invariants under both general coordinate and Lorentz transformations.  In particular, the combination $d^4 V$ in equation (\ref{eq:dV}) is invariant under both coordinate and Lorentz transformations, and thus provides an appropriate volume element for the integration of appropriate field invariants.  If we were dealing with an actual gauge theory, the appropriate kinetic term for the spin connection would be the curvature  squared, but in the case of gravity the situation is slightly different.  In fact, in this case the spin connection is not an independent field, but is determined instead by the vierbein. The only scalar invariant under coordinate transformations and local transformations which contains up to two derivatives of the vierbein is the Ricci scalar,
\begin{equation}
	\mathcal{R}\equiv e^{\mu\,a} e^{\nu\, b} R_{\mu\nu a b}.
\end{equation}
Recall that $\nabla_\mu e^\nu{}_a$ vanishes by construction.

\subsection{Broken Lorentz Symmetry}
\label{sec:Broken Lorentz Symmetry}

The extension of this formalism to theories with broken Lorentz invariance is relatively straight-forward, and parallels the standard construction in flat spacetime. We begin by constructing the most general theory invariant under (linearly realized) local transformations in a  Lorentz subgroup $H$ and general coordinate transformations, and then we show that, by introducing Goldstone bosons, the theory can be made explicitly invariant under the full (non-linearly realized) Lorentz group.

\subsubsection{Unitary Gauge}

Let us first postulate the existence of a vierbein  
$e_\mu{}^a$ that transforms linearly under local transformations in a subgroup of the Lorentz group,\begin{equation}\label{eq:H e}
	h(x): e_\mu{}^a(x)\mapsto e'_\mu{}^a(x)=
	\Lambda^a{}_b(h) \, e_\mu{}^b, \quad    h(x)\in H\subset L^\uparrow_+.
\end{equation}
It is the existence of an orthonormal frame in any spacetime with Minkowski signature what forces us to introduce the vierbein in such a representation.  Given this vierbein, we \emph{define} the spacetime metric to be
\begin{equation}
	g_{\mu\nu}\equiv e_\mu{}^a e_\nu{}^b \eta_{ab}.
\end{equation}
It follows then from the definition of the metric that the vierbein forms a set of orthonormal vectors, as in equation (\ref{eq:orthonormality}), 
and that the volume element (\ref{eq:dV}) is invariant both under general coordinate and Lorentz transformations.

In order to construct derivatives that transform covariantly under local transformations in  $H$, we need to postulate the existence of an appropriate connection $\boldsymbol{\omega}_\mu$. If we want to avoid introducing extraneous ingredients into the gravitational sector, we should construct such a gauge field solely in terms of the vierbein, as in the standard construction.  Inspection of equations (\ref{LLTomega}) and (\ref{eq:omega(e)}) reveals that if we define $\boldsymbol{\omega}_\mu$ by equation (\ref{eq:omega(e)}), under an element of $H$ the connection transforms like
\begin{equation}\label{eq:tilde omega transformation}
	h(x): \boldsymbol{\omega}_\mu\mapsto 
	h\,\boldsymbol{\omega}_\mu \, h^{-1}
	+ h \, \partial_\mu  h^{-1}.
\end{equation}
But as opposed to the original construction in which we demanded invariance under the full Lorentz group, the reduced symmetry in the broken case allows us to introduce additional covariant quantities. In particular, expanding the connection in the basis of broken and unbroken generators, 
\begin{equation}\label{eq:E&D}
	\boldsymbol{\omega}_\mu\equiv 
	i \l( D_{\mu m} \, x^m+E_{\mu i} \, t^i \r)
	\equiv i \l( \boldsymbol{D}_\mu+\boldsymbol{E}_\mu \r),
\end{equation}
it is then easy to verify that $\boldsymbol{D}_\mu$ transforms covariantly (under $H$), while $\boldsymbol{E}_\mu$ transforms like a gauge field, 
\begin{subequations}
\begin{eqnarray}\label{eq:tilde covD transformation}
	h(x)&:&\boldsymbol{D}_\mu(x)\mapsto h\boldsymbol{D}_\mu(x) h^{-1},\\
	&{}&\boldsymbol{E}_\mu(x)\mapsto h\boldsymbol{E}_\mu (x)h^{-1}-i \, h \, \partial_\mu  h^{-1}.
\end{eqnarray}
\end{subequations}
These transformation laws are analogous to those in equations (\ref{eq:Minkowski e D E transf}). The only difference, setting $g=h$ and using equation (\ref{eq:h}),  is that in the latter the Lorentz group acts a transformation in spacetime, which changes the spacetime coordinates of the fields, while here the Lorentz group acts internally, and thus leaves the spacetime dependence of the fields unchanged.
 
The transformation properties of $\boldsymbol{E}_\mu$ allow us to define another covariant derivative of the vierbein, 
${	\bar{\nabla}_\rho e_\mu{}^a=\partial_\rho e_\mu{}^a
	-\Gamma^\nu{}_{\mu\rho}e_\nu{}^a
	-i E_{\rho i} (t^i_4)^a{}_b\,  e_\mu{}^b.
}$
But because $\nabla_\rho e_\mu{}^a=0$, this derivative equals ${-i D_{\nu m} (x^m_4)_a{}^b e^\mu{}_b}$,
and therefore does not yield any additional covariant quantity. Finally, from the connection $\boldsymbol{\omega}_{\mu}$ we  define the curvature (\ref{eq:Riemann}), which under (\ref{eq:H e})  transforms like
\begin{equation}
	h(x) :\boldsymbol{R}_{\mu\nu} \mapsto h\, R_{\mu\nu} \, h^{-1}.
\end{equation}

In order to construct invariants  under both diffeomorphisms and local Lorentz transformations, it is convenient to consider quantities that transform as scalars under diffeomorphisms, and tensors under the unbroken Lorentz subgroup $H$. We thus define, in full analogy with equations (\ref{eq:Minkowski Da}),
\begin{equation}\label{eq:frame components}
 	\boldsymbol{\mathcal{D}}_a \equiv e^{\mu}{}_a \boldsymbol{D}_\mu, \quad
	\boldsymbol{\mathcal{E}}_a \equiv e^{\mu}{}_a \boldsymbol{E}_\mu, \quad
	\boldsymbol{\mathcal{R}}_{a b} \equiv e^\mu{}_a e^\nu{}_b \boldsymbol{R}_{\mu\nu}.
 \end{equation}
The quantities  $\boldsymbol{\mathcal{D}}_a$ and $\boldsymbol{\mathcal{E}}_a$ are the appropriate generalization of the covariant derivatives defined in equation (\ref{eq:Minkowski Da}), since they also transform like in equation (\ref{eq:E D transf}), the only difference being again that here the Lorentz group acts as an internal transformation.   As before, the covariant derivatives of any diffeomorphism scalar $\psi$ that transforms in a representation of the unbroken group with generators $t^i$  are defined by equation (\ref{eq:matter tilde covD}), where $E_{\mu i}$ is now given by equation (\ref{eq:E&D}).  
 
By construction, any term solely built from the  covariant quantities $d^4V$, $\mathcal{D}_{a m}$, $\mathcal{R}_{ab}{}^{c d}$, $\psi$ and $\mathscr{D}_a \psi$, which is invariant under global $H$ transformations is also invariant under local transformations in $H$ and diffeomorphisms.  In particular, because the covariant derivatives $D_{a m}$  defined in (\ref{eq:Minkowski Da}) and the the covariant derivatives in equation (\ref{eq:frame components}) transform in the same way under $H$, the  unbroken symmetries now allow us to write down linear  and quadratic terms for the components of the connection $\boldsymbol{\omega}_\mu$ along the directions of the broken generators, as in equation (\ref{eq:Goldstone invariants}).  In an ordinary gauge theory, the quadratic terms give mass to some gauge bosons, but in our context, because the spin connection depends on derivatives of the vierbein, these quadratic terms cannot be properly considered as mass terms for the graviton.  Since  the spacetime metric is $g^{\mu\nu}=e^\mu{}_a e^{\nu a}$, a graviton mass term should be a quartic polynomial in the vierbein. But the only invariants one can construct from the vierbein $e^\mu{}_a$ are field-independent constants. 

\subsubsection{Manifestly Invariant Formulation}

Let us assume now that we have constructed an $H$ invariant action,
\begin{equation}\label{eq:H inv S}
	S[e,\psi]=S[\Lambda(h)e,\mathscr{R}(h)\psi], \quad h(x)\in H,
\end{equation}
where the functional dependence emphasizes that only  $e$ and  $\psi$ are the ``fundamental" fields of the theory, from which the remaining covariant quantities are constructed, as discussed above.   We show next that by  introducing the corresponding Goldstone bosons in the theory $\gamma\equiv \gamma(\pi_m)$, this symmetry can be extended to the full Lorentz group. To that end, let us  assume that the vierbein $e^\mu{}_a$  transforms in a linear representation of the Lorentz group,  as in  equation (\ref{eq:e transf}), and let us define
\begin{equation}\label{eq:vierbein shorthand}
		\tilde{e}_\mu{}^a \equiv \Lambda^a{}_b(\gamma^{-1}) e_\mu{}^b, 
\end{equation}
where $\tilde{e}^\mu{}_a$ is to be regarded as a shorthand for the expression on the right hand side, and $\gamma$ is a function of the Goldstone bosons defined in equation (\ref{gamma}). Let us postulate that under local Lorentz transformations, $\gamma(\pi)$ transforms as in equation (\ref{eq:transformation}), while under $g(x)\in L^\uparrow_+$,
\begin{equation}\label{eq:psi tilde transf}
	g(x):	\psi\mapsto  \mathscr{R}(h(\pi,g))\psi.
\end{equation}
In that case, it follows from the definition (\ref{eq:vierbein shorthand}) that $\tilde{e}$ transforms analogously,
\begin{equation}\label{eq:shorthand transf}
g(x): \tilde{e}^\mu{}_a\mapsto 
	\Lambda_a{}^b(h(\pi,g)) \, \tilde{e}^\mu{}_b.
\end{equation}

The transformation properties  (\ref{eq:psi tilde transf})  and (\ref{eq:shorthand transf}) and the invariance of the action (\ref{eq:H inv S}) imply  that a theory with
\begin{equation}\label{eq:L inv S}
	\tilde{S}[\gamma,e,\psi]\equiv S[\Lambda(\gamma^{-1})e,\psi]
\end{equation}
is invariant under the full Lorentz group.  In the Lorentz-invariant formulation of the theory in equation (\ref{eq:L inv S}) the action appears to depend on the Goldstone bosons $\gamma(\pi)$. However, inspection of the right hand side of the equation reveals that such a dependence can be removed by the field redefinition (\ref{eq:vierbein shorthand}).  By a ``field redefinition" we mean here a change of variables in the theory, which replaces the combination of two fields $\Lambda(\gamma^{-1})e$ by a single field, which we may call again $e$.   Since the field variables we use do not have any impact on the physical predictions of a theory,  we may thus replace $S[\Lambda(\gamma^{-1})e,\psi]$ by 
$S[e,\psi]$. In this ``unitary gauge" we have effectively set $\gamma=1$, and returned back to the original action in equation (\ref{eq:H inv S}).

It is instructive to show how the introduction of the Goldstone bosons would make the theory manifestly invariant under local transformations. For simplicity, let us just focus on the gravitational sector. As mentioned above, the modified vierbein (\ref{eq:vierbein shorthand}) transforms non-linearly under the action (\ref{eq:e transf}) of the Lorentz group, $g(x)\in L^\uparrow_+$.  When we substitute this modified vierbein into the expression for the spin connection (\ref{eq:omega(e)}) we obtain
\begin{equation}\label{eq:tilde omega}
	\tilde{\boldsymbol{\omega}}_\mu=\gamma^{-1} \l( \partial_\mu + \tilde{\boldsymbol{\omega}}_\mu \r) \gamma,
\end{equation}
which is just the covariant generalization of the Maurer-Cartan form $\gamma^{-1}\partial_\mu \gamma$, and transforms non-linearly under (\ref{eq:e transf}),
\begin{equation}
	g(x): \tilde{\boldsymbol{\omega}}_\mu\mapsto 
	h(\pi,g) \, \boldsymbol{\omega}_\mu \, h^{-1}(\pi,g)
	+ h(\pi,g) \, \partial_\mu h^{-1}(\pi,g),
\end{equation}
with $h(\pi,g)$ defined in equation (\ref{eq:transformation}). Therefore, if we expand this connection in the basis of the Lie algebra,
\begin{equation}\label{eq:tilde D}
	\tilde{\boldsymbol{\omega}}_\mu\equiv 
	  i\left[\tilde{D}_{\mu m} \, x^m+\tilde{E}_{\mu i} \, t^i\right],
\end{equation}
we obtain covariant derivatives ${\tilde{\boldsymbol{\mathcal{D}}}_a\equiv \tilde{e}^\mu{}_a \tilde{\boldsymbol{D}}_\mu}$ and gauge fields ${\tilde{\boldsymbol{\mathcal{E}}}_a\equiv \tilde{e}^\mu{}_a \tilde{\boldsymbol{E}}_\mu}$  that transform like in equations (\ref{eq:E D transf}), but with $x'=x$. The curvature tensor $\tilde{\boldsymbol{R}}_{\mu\nu}$ associated with $\tilde{\boldsymbol{\omega}}_\mu$ is in fact given by 
\begin{equation}\label{eq:tilde R}
	\tilde{\boldsymbol{R}}_{\mu\nu}=\gamma^{-1} \boldsymbol{R}_{\mu\nu}\gamma,
\end{equation}
 where $\boldsymbol{R}_{\mu\nu}$ is the curvature tensor associated with the spin connection $\boldsymbol{\omega}_\mu$, derived itself from $e_\mu{}^a$.  Under the action of elements $g(x)\in L^\uparrow_+$ on the vierbein (\ref{eq:e transf}), this curvature transforms non-linearly too, 
\begin{equation}
	g(x) : \tilde{\boldsymbol{R}}_{\mu\nu} \mapsto h(\pi,g)\, \tilde{\boldsymbol{R}}_{\mu\nu} \, h^{-1}(\pi,g).
\end{equation}

It is thus clear from the transformation properties of these new quantities that if the original action $S$ is invariant under $H$, the new action $\tilde{S}$ defined in equation (\ref{eq:L inv S}) will be invariant under $L^\uparrow_+$. In fact, we could have reversed the whole construction. We could have started by defining a modified vierbein $\tilde{e}_\mu{}^a$, a modified covariant derivative $\tilde{\boldsymbol{D}}_\mu$ and a modified curvature tensor $\tilde{\boldsymbol{R}}_{\mu\nu}$ according to equations (\ref{eq:vierbein shorthand}), (\ref{eq:tilde D}) and (\ref{eq:tilde R}).  Then, any invariant action under $H$, solely constructed out of these ingredients would have been automatically and manifestly invariant under $L^\uparrow_+$. 

\section{Unbroken Rotations}
\label{sec:Unbroken rotations}

We turn now our attention to cases in which the unbroken group is the rotation group, $H=SO(3)$, which is the maximal compact subgroup of $L^\uparrow_+$. This pattern of symmetry breaking is analogous to the spontaneous breaking of chiral invariance in the two quark model. In the latter,  the chiral symmetry of QCD with two massless quarks, ${SU(2)_L\times SU(2)_R}$, is broken down to the isospin  subgroup $SU(2)$, while in the former, the Lorentz group $SO(1,3)\sim SU(2)\times SU(2)$ is broken down to the diagonal subgroup of rotations $SO(3)\sim SU(2)$. Hence, the construction of rotationally invariant Lagrangians with broken Lorentz invariance is formally analogous to the construction of isospin invariant Lagrangians with broken chiral symmetry. 

As in the two-quark model, the case for unbroken rotations can be motivated phenomenologically.  If rotations were broken, we would expect the expansion of the universe to be anisotropic, in conflict with observations, which are consistent with a nearly isotropic cosmic expansion  all the way from the initial stages of inflation.    Our main goal here however is not to consider the phenomenology of theories with unbroken rotations, as this has been already extensively studied, but simply  to illustrate how our formalism applies to theories with gravity. We shall see in particular  how in this case our construction directly leads to the well-known Einstein-aether theories, which we show to be the most general class of theories in which rotations remain unbroken.

\subsection{Coset Construction}

In order to build the most general theory in which the rotation group remains unbroken, let us assume first that spacetime is flat, as in Section \ref{sec:coset construction}.  In the case at hand, then, the generators of the unbroken group are the generators of rotations $J_i$, and the remaining ``broken" generators are the boosts $K_m$.  Therefore, the theory contains three Goldstone bosons $\pi_m$. Of particular relevance are the transformation properties of these Goldstone bosons under rotations. For an infinitesimal rotation $t=\omega^i J_i$,  equations (\ref{eq:Cartan})  and (\ref{eq:R}) lead to
\begin{equation}
	t: \pi_m\mapsto \pi'_m=\pi_m+ \l(\omega \times \pi\r)_m.
\end{equation}
In addition, since $P K^m P^{-1} = -K^m$ and $T K^m T^{-1} = -K^m$ we have, from (\ref{eq:gamma transf V})  that ${\pi_m \to -\pi_m}$ under parity and time reversal. Therefore, the set of Goldstone bosons transform like a 3-vector under spatial rotations.  These are analogous  to the pions of spontaneously broken chiral invariance.

The restriction of the four-vector representation $\Lambda(g)$ to the subgroup of rotations $H$ is reducible, ${\bf 4}={\bf 1}\oplus{\bf 3}$, so the tensor product representation of the rotation group in equation (\ref{eq:tensor})  is also reducible, 
\begin{equation}
	({\bf 1}\oplus{\bf 3})\otimes {\bf 3}=
	\bf{3}\oplus {\bf 1}\oplus {\bf 3}\oplus {\bf 5}.
\end{equation}
(The different representations of the rotation group are labeled by their dimension. The dimension $N$ of the representation is $N=2S+1$, where $S$ is the spin of the representation.) More precisely, the covariant derivative
\begin{equation} \label{covderrot}
	\mathcal{D}_m\equiv \mathcal{D}_{0m} 
\end{equation}
transforms like a  spatial vector under rotations (spin one, ${\bf 3}$),  while $D_{mn}$ transforms in the tensor product representation of rotations ${\bf 3}\otimes {\bf 3}$. Defining 
\begin{equation} \label{Dmnrot}
	\mathcal{D}_{mn}=\frac{1}{3}\, \vphi \, \delta_{mn} + a_{mn} + s_{mn},
\end{equation}
with $a$ antisymmetric and $s$ symmetric and traceless, leads to a  scalar $\vphi$ (spin zero, ${\bf 1}$), a  vector $a_{mn}\equiv \epsilon_{mnp}a^p$ (spin one, ${\bf 3}$),  and a  traceless symmetric tensor $s_{mn}$ (spin two,  ${\bf 5}$).
Therefore, the most general Lagrangian density at most quadratic in the covariant derivatives, and invariant under the \emph{full} Lorentz group is 
\begin{equation}\label{eq:rotation invariants}
	\mathcal{L}_\pi = \frac{1}{2} \l( F_\vphi\, \vphi^2+F_D  \mathcal{D}_m \mathcal{D}^m+F_a \, a_{mn} a^{mn}+F_s \, s_{mn}s^{mn} \r) ,
\end{equation}
where indices are raised with the (inverse)  metric of Euclidean space, $\delta^{mn}$. Note that we have omitted a linear term proportional to $\varphi$, and the parity-violating expression $\epsilon_{mnp}\, a^{mn} \mathcal{D}^p$ in the Lagrangian. As we show below, these terms are just  total derivatives.

Let us now address  the new ingredients that gravity introduces into the theory. As we discussed in Section \ref{sec:Broken Lorentz Symmetry},  in a generally covariant theory we may choose to work in unitary gauge, in which the Goldstone bosons identically vanish. In this gauge, the covariant derivatives  $\mathcal{D}_{a m}$ defined  above simply reduce to the spin connection along the appropriate generators, as in equations (\ref{eq:E&D}). Therefore, using the explicit form of the rotation generators in the fundamental representation, and  ${\mathrm{tr} (x^m_{(4)} \cdot x^n_{(4)})=-2 \delta^{mn}}$, we find 
\begin{equation}\label{eq:covD spin}
	\mathcal{D}_{m}=\omega_{0m0}, \quad \mathcal{D}_{mn}=\omega_{mn0}.
\end{equation}
Recall that there are three broken generators which transform like vectors under rotations, which we label by $m, n$, and that the derivatives defined in  equations   (\ref{eq:frame components}) transform in the same way as the covariant derivatives defined in equation (\ref{eq:Minkowski Da}), with $x'=x$. Therefore, the Lagrangian (\ref{eq:rotation invariants}) already contains all the rotationally invariant terms constructed from the undifferentiated spin connection.

To complete the most general gravitational action invariant under general coordinate and local Lorentz transformations, with at most two derivatives acting on the vierbein, we just need to add all invariant terms that can be constructed from the curvature alone. Without loss of generality, we may restrict ourselves to the components of the Riemann tensor in an orthonormal frame,  $\mathcal{R}_{ab}{}^{cd}$. Then,  indices along spatial direction transform like vectors, while  indices along the time direction transform like scalars under rotations. Most of the invariants one can construct out of the Riemann tensor vanish because of antisymmetry. For instance, the term $\mathcal{R}_0{}^{mnp} \epsilon_{mnp}$ is identically zero because of the antisymmetry of the curvature tensor in the last three indices. In addition, the identity ${[\nabla_\mu,\nabla_\nu]A^\rho=R_{\mu\nu}{}^\rho{}_\sigma A^\sigma}$, in an orthonormal frame and up to boundary terms, implies the relation
\begin{equation}
	\int d^4V \left[\mathcal{R}_{0m}{}^{0m}-\mathcal{D}_{mn}\mathcal{D}^{mn}+(\mathcal{D}_m{}^m)^2\right]=0,
\end{equation}
which can be used to eliminate a scalar term proportional to 	$\mathcal{R}_{0m}{}^{0m}$ from the action.   As we mentioned earlier a term linear in the covariant derivative, $\varphi\equiv \mathcal{D}_m{}^m$, is a total derivative, since from equations (\ref{eq:covD e}) and (\ref{eq:covD e zero})
\begin{equation}
	\omega_{m0}{}^m=\partial_\mu e^\mu{}_0+\Gamma^\mu{}_{\nu\mu}e^\nu{}_0
	=\frac{1}{\det e}\partial_\mu(\det e \,  e^\mu{}_0).
\end{equation}
Similarly, one can show that $\epsilon_{mnp}\, a^{mn} \mathcal{D}^p$ is a total derivative too, since the latter equals ${\epsilon^{mnpq}\nabla_m A_n \nabla_p A_q}$, for $A_m=\delta_m{}^0$. We therefore conclude that the most general diffeomorphism invariant action invariant under local rotations is
\begin{equation}\label{eq:rotational invariant action}
	S=\frac{M_P^2}{2} \int d^4 V \left[\mathcal{R}+ \mathcal{L_\pi}\right]+S_M,
\end{equation}
where $\mathcal{R}\equiv \mathcal{R}_{ab}{}^{ab}$ is the Ricci scalar,  the ``Goldstone" Lagrangian $\mathcal{L_\pi}$ is given by equation (\ref{eq:rotation invariants}), and $S_M$ denotes the matter action.  Tests of the equivalence principle  \cite{Will:2005va} and constraints on Lorentz-violating couplings in the standard model \cite{Kostelecky:2008ts} suggest that any Lorentz-violating term in the matter action $S_M$ is very small. Hence, for phenomenological reasons, we assume that the breaking of Lorentz invariance is restricted to the gravitational sector. Therefore, $S_M$ is taken to be invariant under Lorentz transformations, and the action  (\ref{eq:rotational invariant action}) defines a metric theory of gravity.   

\subsection{The Einstein-aether}

For unbroken rotations, the matrix $\gamma$ that we introduced in Section \ref{sec:coset construction} is a boost, ${\gamma =\exp(i \pi_m K^m)}$. Hence, instead of characterizing the Goldstone bosons by the set of three scalars $\pi_m$, we may simply describe them by the transformation matrix $\Lambda^a{}_0$ of the boost itself. The latter has four components,
\begin{equation}\label{eq:boost}
	u^a\equiv \Lambda^a{}_0,
\end{equation}
but not all of them are independent, because Lorentz transformations preserve the Minkowski metric. In particular, the vector field $u_a$ has unit norm
\begin{equation}\label{eq:unit norm}
	u_a u^a\equiv \eta_{ab} \Lambda^a{}_0   \Lambda^b{}_0=\eta_{00}=-1.
\end{equation}
In the conventional approach to the formulation of the most general theory in which rotations remain unbroken, one would solve the constraint (\ref{eq:unit norm}) by introducing an appropriate set of three parameters, and then identify their transformation properties under the Lorentz group \cite{Graesser:2005bg}. One would then proceed to define covariant derivatives of these parameters, and use them to construct the most general theory compatible with the unbroken symmetry, just as we did.

In this case however, a simpler approach leads to the same general theory, but avoids introducing coset parametrizations and covariant derivatives altogether.  Since the Lorentz transformation of a boost can be described by a the vector field (\ref{eq:boost}), one may simply expect that the problem of constructing the most low-energy effective theory in which the rotation group remains unbroken just reduces to the problem of writing down the most general diffeomorphism invariant theory with the least numbers of derivatives acting on a unit norm vector field. This was precisely the problem that  Jacobson and Mattingly studied in \cite{Jacobson:2000xp}, which resulted in what they called the ``Einstein-aether". The most general action in this class of theories is
\begin{multline}\label{eq:Einstein aether}
	S=\frac{M_G^2}{2}\int d^4V \Big[\mathcal{R} - c_1 \nabla_a u_b \nabla^a u^b - c_2 (\nabla_a u^a)^2
	 - c_3 \nabla_a u_b \nabla^b u^a+\\
	{}+ c_4 u^a u^b \nabla_a u_c \nabla_b u^c+\lambda(u_a u^a+1)\Big],
\end{multline}
where the parameters $c_i$ are constant,  and we have written down all the components of the ``aether"  vector field $u^\mu$ in an orthonormal frame,  ${u^a\equiv e_\mu{}^a u_\mu}$, with covariant derivatives given by
\begin{equation}\label{eq:covD vector}
	\nabla_a u^b\equiv e^\mu{}_a \l( \partial_\mu u^b+\omega_\mu{}^b{}_c u^c \r).
\end{equation}
The  constraint  $u^a u_a= - 1$ on the norm of the field is enforced  by the Lagrange multiplier $\lambda$. Hence, the action (\ref{eq:Einstein aether})  is analogous to the linear $\sigma$-model  in which chiral symmetry breaking was originally studied. In this formulation, the Lorentz group acts linearly on the vector field $u^a$, and, as we shall see, the fixed-norm constraint can be understood as  limit in which the potential responsible for Lorentz symmetry breaking is infinitely steep around its minimum.

To establish the connection between the Einstein-aether (\ref{eq:Einstein aether})  and the rotationally invariant action (\ref{eq:rotational invariant action}), we simply need to impose unitary gauge. We can solve the unit norm constraint in (\ref{eq:Einstein aether}) by  expressing the  vector field $u^a$  as a Lorentz transformation acting on an appropriately chosen vector $\tilde{u}^a$, 
\begin{equation}
	u^a = \Lambda^a{}_b (\pi) \tilde{u}^b, \quad \text{with} \quad \tilde{u}^a=\delta^a{}_0,
\end{equation}
which  is just a restatement of equation (\ref{eq:boost}). Then, invariance under local Lorentz transformations implies that the aether action (\ref{eq:Einstein aether}) can be equally thought of as a functional of $\tilde{u}^b$ and the transformed vierbein $ \tilde{e}_\mu{}^a = (\Lambda^{-1}(\pi))^a{}_b \, e_\mu{}^b$. 
If we now redefine the vierbein field, $\tilde{e}_\mu{}^b\to e_\mu{}^a$, the Goldstone bosons $\pi$ disappear from the action, and we are left with the  theory in  unitary gauge.  In this gauge the vierbein is arbitrary, but (dropping the tildes) we can assume that $u^a =\delta^a{}_0$.  In that case equation (\ref{eq:covD vector})  gives in addition $\nabla_a u_b =\omega_{ab0}$, which, when substituted into the Einstein-aether action (\ref{eq:Einstein aether}) precisely yields  the action (\ref{eq:rotational invariant action}). The corresponding  parameters $M_P$ and $F_i$  are expressed in terms of five linearly independent combinations of aether parameters,
\be \label{eq:dictionary}
	M_P = M_G, \quad F_\vphi = -\frac{1}{3}(c_1+3c_2+c_3), \quad
	 F_D = c_1+c_4,  \quad F_a = c_3 - c_1, \quad F_s = -(c_1+c_3), 
\ee
and, therefore, the Einstein-aether is the most general low-energy theory in which the rotation group remains unbroken. The correspondence (\ref{eq:dictionary})  also explains then why these particular combinations of the Einstein-aether parameters enter the predictions of the theory. In our language, they map  into the different irreducible representations in which one can classify the covariant derivatives of the Goldstone bosons.  The phenomenology of Einstein-aether theories is nicely reviewed in \cite{Jacobson:2008aj}.

\subsection{General Vector Field Models}\label{vector example}

In Einstein-aether theories, Lorentz invariance is broken  because the vector field $u^a$ develops a time-like vacuum expectation value. In this context, it is then natural to consider generic vector field theories in which a vector field develops a non-zero expectation value, and to study how the latter reduce to the Einstein-aether in the limit of low energies.  This will also help us to illustrate our formalism in cases in which the spectrum of excitations contains a massive field, and how the latter disappears from the low-energy predictions of the theory. 

The most general low energy effective action for a vector field non-minimally coupled to gravity which contains at most two derivatives and is invariant under local Lorentz transformations and general coordinate transformations reads
\ba \la{Riccardos-lagr}
S =  \frac{1}{2}\int d^{4}V \hspace{-.35cm}&&\bigg[M_{G}^{2} \mathcal{R}+\fr{\alpha}{2} \, F_{ab} F^{ab}  + \beta \l( \nabla_{a} A^{a} \r)^2 + \, \beta_4 \mathcal{R} A_{a} A^{a} + \beta_5 \mathcal{R}_{ab} A^{a} A^{b}+ \\
&& \qquad + \fr{A^{a} A^{b}}{\Lambda^2} \l( \alpha_1 \, \nabla_{a} A_{c} \nabla_{b} A^{c} + \alpha_2 \, \nabla_{c} A_{a} \nabla^{c} A_{b} + \alpha_3 \, \nabla_{a} A_{b} \nabla_{c} A^{c} \r)+ \nonumber\\
&& \qquad\qquad\quad    \left.   + \gamma \, \fr{A^{a} A^{b}A^{c} A^{d}}{\Lambda^4} \nabla_{a} A_{b} \nabla_{c} A_{d} + \delta_1   A_b A^b \nabla_{a} A^{a} - \Lambda^4 \, V  \right] .\nonumber 
\ea
Here, $F_{ab} \equiv \partial_{a} A_{b} - \partial_{b} A_{a}$, $A^a$ are the components of the vector field in  an arbitrary orthonormal frame, and the various coefficients $\alpha$, $\alpha_i$, $\beta$, $\beta_i$, $\gamma$, $\delta_{1}$ and $V$ should be regarded as arbitrary (dimensionless) functions of $A_{a}A^{a} /\Lambda^2$. Finally, $M_{G}$ and $\Lambda$ are the two characteristic energy scales of the effective theory, which is valid at energies $E \ll \textrm{min}(\Lambda, M_{G})$. In order to generate spontaneous breaking of Lorentz symmetry down to rotations we assume, without loss of generality, that the potential $V$ is minimized by field configurations with $A_a A^a = -\Lambda^2$. Other low energy terms that do not appear in the expression (\ref{Riccardos-lagr}) can be reduced to linear combinations of the terms above after integrations by parts. An action very similar to (\ref{Riccardos-lagr}) has been already considered in \cite{Gripaios:2004ms}, though the latter did not include the terms proportional to $\beta_4$ and $\delta_1$, and  all the other couplings were assumed to be constants rather than  arbitrary functions of $A^a$. Models involving fewer terms have been studied  for instance in \cite{Bluhm:2004ep,Bailey:2006fd,Bluhm:2007bd,Kostelecky:2009zr} under the name of ``bumblebee models," and in \cite{ArmendarizPicon:2009ai} under the name of ``unleashed aether models."

In order to make contact with the formalism developed in the previous sections, we shall parametrize again the vector field  as a Lorentz transformation acting on 
\begin{equation}\label{eq:A unitary}
	A^a (x) = \delta^a_0 \left(\Lambda +\sigma(x)\right),
\end{equation}
where the field $\sigma$ is just a singlet under rotations.  This is the same we did for the aether, the only difference being that there the fixed-norm constraint forced the field $\sigma$ to vanish.  As before, invariance under local Lorentz transformations then implies that the vector field can be taken to be given by (\ref{eq:A unitary}). In this unitary gauge,  the covariant derivative of $A^a$ is
\begin{equation}
	\nabla_{a}A^{b}=\delta^{b}_{0}(e^{\mu}{}_{a}\partial_{\mu}\sigma)+\eta^{bm}(\Lambda+\sigma)\, \mathcal{D}_{am},
\end{equation}
where we have used equations (\ref{eq:covD spin}). Thus,  the action (\ref{Riccardos-lagr}) can be expressed in terms of rotationally invariant operators that solely involve $\mathcal{R}_{ab}{}^{cd}$, $\mathcal{D}_{a m}$, the scalar $\sigma$ and its covariant derivative  $\mathscr{D}_a \sigma = e^\mu{}_a \d_\mu \sigma$.

It shall prove to be useful to expand the action (\ref{Riccardos-lagr}) in powers of $\sigma.$ To quadratic order, and to leading order in derivatives, this results is
\begin{eqnarray}
	&\displaystyle S = \frac{1}{2}\int d^{4}V \bigg[ (M^2_G - \bar{\beta}_4 \Lambda^2) \mathcal{R}+  \Lambda^2 \l (\bar{\beta} + \frac{2 \bar{\beta}_5}{3} \r) \vphi^2 + \Lambda^2 (\bar{\alpha}_1 - \bar{\alpha}) \mathcal{D}_m \mathcal{D}^m+&  \label{eq:vector unitary}  \\
	  & \displaystyle+ \Lambda^2 (2 \bar{\alpha}+ \bar{\beta}_5) \, a_{mn} a^{mn} - \Lambda^2 \bar{\beta}_5 \, s_{mn} s^{mn}
	+\sigma(- 2 \bar{\delta}_1 \Lambda^2 \varphi+ \cdots)
	+\sigma^2(-2 \bar{V}'' \Lambda^2+\cdots)
	+\mathcal{O}(\sigma^3)\bigg],& \nonumber
\end{eqnarray}
where the dots stands for the subleading terms in the derivative expansion and $\bar{V}''$ denotes the second derivative of the potential function with respect to its argument, evaluated at its minimum, where $A_a A^a = - \Lambda^2$. Similarly, $\bar{\alpha},\bar{\beta}, \bar{\beta}_4, \bar{\beta}_5, \bar{\alpha}_1$ and $\bar{\delta}_1$ stand for the values of the couplings at the minimum of the potential. Apart from the additional rotationally invariant terms involving the field $\sigma$, the action (\ref{eq:vector unitary}) has manifestly the form (\ref{eq:rotational invariant action}) with ${M_{P}^{2}\equiv(1-\bar{\beta}_{4})M_{G}^{2}}$. 

We study the spectrum  of this class of theories in Appendix \ref{app: dispersion}. Their scalar sector consists of a massless excitation, one of the Goldstone bosons, and a massive field, whose mass is linear in  $\bar{V}''$.  We show in the appendix that in the low-momentum limit, the field $\sigma$ has a vanishing matrix element  between  the massless scalar particle and the vacuum,
\begin{equation} \la{matrix-limit}
	\lim_{p\to 0} \, \langle m=0  | \sigma(p)|0\rangle=0.
\end{equation}
Hence, if we are interested in low momenta and massless excitations, the field $\sigma$ can be simply integrated out. At tree level, this can be easily done by solving the classical equations of motion to express $\sigma$ in terms of the covariant derivatives $\mathcal{D}_{a m}$. From (\ref{eq:vector unitary}), we see that to lowest order in derivatives the result is completely determined by the two terms proportional to $\sigma^2$ and $\sigma\varphi$. Thus, solving the corresponding linear equation,
\be \la{sigma sol}
\sigma = - \frac{\bar{\delta}_1^2}{2 \bar{V}''}\varphi + \mathcal{O} (\d^2/\Lambda), 
\ee
and  plugging back into the action (\ref{eq:vector unitary}) we get, to leading order in derivatives, 
\ba\label{eq:integrated out}
S &=& \frac{1}{2} \int d^4 V  \bigg[ (M^2_G - \bar{\beta}_4 \Lambda^2) \mathcal{R} + \Lambda^2 \l (\bar{\beta} + \frac{\bar{\delta}_1^2}{2 \bar{V}''}+ \frac{2 \bar{\beta}_5}{3} \r) \vphi^2 + \Lambda^2 (\bar{\alpha}_1 - \bar{\alpha}) \mathcal{D}_m \mathcal{D}^m+ \nonumber  \\
&& \qquad \qquad \qquad \qquad\qquad \qquad \qquad \qquad  + \Lambda^2 (2 \bar{\alpha} + \bar{\beta}_5) \, a_{mn} a^{mn} - \Lambda^2 \bar{\beta}_5 \, s_{mn} s^{mn} \bigg]. 
\ea
As expected the low energy action (\ref{eq:integrated out}) has the form of (\ref{eq:rotational invariant action}). Integrating out the field sigma has simply renormalized the coefficients of the low energy theory, which are now given by 
\ba
&\displaystyle M_P^2 = M_G^2 - \bar{\beta}_4 \Lambda^2, \qquad F_\vphi = \l(\bar{\beta} + \frac{\bar{\delta}_1^2}{2 \bar{V}''} + \frac{2 \bar{\beta}_5}{3} \r) \frac{\Lambda^2}{M_P^2}, \qquad F_D = (\bar{\alpha}_1 - \bar{\alpha}) \frac{\Lambda^2}{M_P^2},& \nonumber \\
&\displaystyle F_a =  (2 \bar{\alpha} + \bar{\beta}_5) \frac{\Lambda^2}{M_P^2}, \qquad F_s = - \bar{\beta}_5 \frac{\Lambda^2}{M_P^2}. & \la{mapping}
\ea
By combining these relations with equations (\ref{eq:dictionary}), one can easily derive the dispersion relations and residues of the massless excitations in the model (\ref{Riccardos-lagr}) from the known aether theory results \cite{Jacobson:2008aj}. Equations (\ref{mapping}) show from the very beginning that the couplings $\gamma, \alpha_2$ and $\alpha_3$ will not enter the low-energy phenomenology. A ``brute force'' calculation based on the action (\ref{Riccardos-lagr}) tends to obscure this fact, as shown explicitly in Appendix \ref{app: dispersion}, although the final results are of course identical.

Alternatively, if we are interested only in the low energy phenomenology of the theory, we can choose to drop the field $\sigma$ from the onset, as massive excitations will not give any observable contribution at low energies \cite{Appelquist:1974tg}. In the limit $\bar{V}''\to \infty$ where the massive mode becomes infinitely heavy, the potential may be replaced by a fixed-norm constraint, as in Einstein-aether theories. In fact, when $\bar{V}''\to \infty$, equation (\ref{sigma sol}) implies that $\sigma$ can be simply set to zero, and the general class of vector field models described by (\ref{Riccardos-lagr}) directly reduces to the Einstein-aether. After introducing a rescaled vector $A^a \equiv \Lambda u^a$ and integrating some terms by parts, the coefficients $c_i$  in (\ref{eq:Einstein aether}) can be easily mapped onto the couplings in (\ref{Riccardos-lagr}) as follows:
\ba
&\displaystyle \alpha = - c_1 \frac{M_G^2}{\Lambda^2},\qquad \beta = -(c_1+c_2+c_3) \frac{M_G^2}{\Lambda^2}, \qquad \beta_5 = (c_1+c_3) \frac{M_G^2}{\Lambda^2}, \qquad \alpha_1 = c_4 \frac{M_G^2}{\Lambda^2},& \nonumber \\
&\alpha_2 = \alpha_3 = \beta_4 = \gamma = \delta_1 = 0.& \la{vector-tensorTOeinstein-aether}
\ea
Once again, equations (\ref{vector-tensorTOeinstein-aether}) can be easily combined with  the known  Einstein-aether results  \cite{Jacobson:2008aj} to immediately obtain the dispersion relations and the residues for the massless propagating modes in the specific model (\ref{Riccardos-lagr}).

\section{Summary and Conclusions}
\label{sec:Conclusions}

In this article we have generalized the effective Lagrangian construction of Callan, Coleman, Wess and Zumino to the Lorentz group. In flat spacetime, the Lorentz group is a global symmetry, and its breaking implies the existence of  Goldstone bosons, one for each broken Lorentz generator. The broken global symmetry is not lost, and is realized non-linearly in the transformation properties of these Goldstone bosons and the matter fields of the theory. Because the Lorentz group is a spacetime symmetry, the Goldstone bosons transform non-trivially under the Lorentz group, and can be classified in linear representations of the unbroken subgroup.  The same non-linearly realized global symmetry prevents the Goldstone bosons from entering the Lagrangian undifferentiated, which allows us to identify them as massless excitations.  Because spacetime derivatives transform non-trivially under the Lorentz group, the covariant derivatives of Goldstone bosons typically furnish reducible representations of the unbroken Lorentz subgroup. The Lorentz group does not seem to be broken in the standard model sector, so any eventual breaking of this symmetry must be confined to a hidden sector of the theory.  In that respect, phenomenologically realistic  theories must resemble models of gravity-mediated supersymmetry breaking \cite{Nilles:1982ik,Chamseddine:1982jx,Barbieri:1982eh}. In both cases, a spacetime symmetry is broken in a hidden sector, the breaking is communicated to the standard model by the gravitational interactions, and, for phenomenological reasons,  the symmetry breaking scale has to be sufficiently low. 

Given an internal symmetry group, one always  has a choice to make it global or local. But in the case of the Lorentz group this choice does not seem to exist. Any generally covariant theory that contains spinor fields, such as the standard model coupled to general relativity,  requires that Lorentz transformations be an internal local symmetry, very much like a group of internal gauge symmetries. We have therefore extended the construction of actions in which global Lorentz invariance is broken to generally covariant formulations in which the group of local Lorentz transformations is non-linearly realized on the  fields of the theory, which at the very least must contain the covariant derivatives of the Goldstone bosons and the vierbein, which describes the gravitational field.  But in this case, since the Lorentz group is a local symmetry, it is possible and simpler to work in a formulation  in which the Goldstone bosons are absent, and Lorentz symmetry is explicitly broken. In this ``unitary gauge," the theory remains generally covariant, but Lorentz symmetry is lost.   Even though the lost invariance under the Lorentz group can always be restored by introducing the appropriate Goldstone bosons,  this restored symmetry is merely an artifact. 

Generally covariant theories with broken Lorentz invariance differ significantly from their fully symmetric counterparts.  In unitary gauge for instance, the covariant derivatives of the Goldstone bosons that the unbroken symmetry allows us to write down simply become the spin connection along the broken generators. This is just the Higgs mechanism. But in a generally covariant theory without extraneous additional fields, this connection is expressed in terms of the vierbein, so these terms actually represent kinetic terms for some of its components. Thus, instead of a massive theory of gravity, when Lorentz invariance is broken we obtain a theory with additional massless excitations (in Minkowski spacetime), which we can interpret as extra graviton polarizations in unitary gauge, or  simply as the Goldstone bosons  of the theory in general.

We have illustrated these issues for cases in which the rotation group remains unbroken. In particular, we have rigorously shown that the most general low-energy effective theory with unbroken spatial rotations is the Einstein-aether, and how generic vector field theories reduce to the latter at low energies. 

The construction of  low-energy effective  theories that we have described here provides us with a tool to explore Lorentz symmetry breaking systematically and in a model-independent way. It identifies first how the Lorentz group acts on the field of the theory, it removes the clutter of particular models by focusing on the relevant fields at low energies, and it uniquely  enumerates  all the invariants under the unbroken symmetries.  

\acknowledgments

The  work of CAP and RP is supported in part by the National Science Foundation under Grant No. PHY-0855523. The work of ADT is supported by a UNAM postdoctoral fellowship.

\appendix

\section{Dispersion Relations for Vector-Tensor Effective Theories} \la{app: dispersion}

In this appendix we study the spectrum of excitations  in  the vector-tensor theories introduced in Section \ref{vector example}, in which Lorentz symmetry is spontaneously broken down to rotations. Although such a study is usually carried out in the standard metric formulation (see for example \cite{Gripaios:2004ms}), in what follows we adopt instead the vierbein formulation, which is the one we employ in  the main body of this paper. 

\subsection{Perturbations}
Our starting point is  the action (\ref{Riccardos-lagr}), which is a functional of the vierbein $e_\mu{}^a$ and the vector field $A^a$, and describes the behavior of both light and heavy modes. Perturbations of the vierbein around the Minkowski solution $e_\mu{}^{a} = \delta_\mu{}^{a}$ can be decomposed into scalars, vectors and tensors under spatial  rotations as follows:
\begin{subequations}\label{general-pert}
\begin{eqnarray}
\delta e_0{}^0 &=& \phi, \\
\delta e_0{}^i &=&  \partial_i B + S_i, \\
\delta e_i{}^0 &=& - \partial_i C - T_i, \\
\delta e_i{}^j &=& - \delta_{ij} \psi + \partial_i \partial_j E + \epsilon_{ijk} \partial^k D - \partial_{(i} F_{j)} + \epsilon_{ijk} W^k +\frac{1}{2} h_{ij}.
\end{eqnarray}
\end{subequations}
In this decomposition $\phi, B, C, \psi, E, D$ are scalars, $S_i, T_i, F_i, W_i$ are transverse vectors, ${\partial_i S^i=\cdots=\partial_i W^i=0}$, and $h_{ij}$ is a transverse and traceless tensor, $h_i{}^i=\partial_i h^{ij}=0$. Here, $i=1,2,3$ labels spatial indices, which  we raise and lower with the flat metric $\delta^{ij}$.

Scalars, vectors and tensors  transform in different irreducible representations of the rotation group and therefore do not couple from each other in the free theory. As we show in Section \ref{vector example}, no matter what the spacetime background is, we can always use invariance under local boosts to impose the ``unitary gauge" condition (\ref{eq:A unitary}), namely
\begin{equation} \la{eq:A unitary-2}
	A^a (x) = \delta^a_0 \left(\Lambda +\sigma(x)\right).
\end{equation}
The field $\sigma$ is a scalar under rotations.

\subsubsection*{Gauge fixing}
At this point, not all the scalars and vectors in equations (\ref{general-pert}) and (\ref{eq:A unitary-2}) describe independent degrees of freedom, because of the residual gauge invariance associated with general coordinate transformations and the unbroken group of local rotations.
In fact, under infinitesimal coordinate transformations ($x^\mu\to x^\mu+\xi^\mu$) and local Lorentz rotations ($e_i{}^\mu \to e_i{}^\mu + \omega^k \epsilon^{ij}{}_k e_j{}^\mu$) the fluctuations of the vierbein around a Minkowski background (\ref{general-pert}) transform in the following way:
\begin{subequations}\label{general-transformation}
\begin{eqnarray}
\delta e_0{}^0 &\to& \delta e_0{}^0-\partial_t \xi^0, \\
\delta e_0{}^i &\to&  \delta e_0{}^i-\partial_t\partial^i \xi-\partial_t\xi^i_T,\\
\delta e_i{}^0 &\to&   \delta e_i{}^0-\partial_i \xi^0, \\
\delta e_i{}^j &\to & \delta e_i{}^j -\partial_i \partial^j\xi-\partial_i \xi_T^j+\epsilon_i{}^{jk} \partial_k \omega +\epsilon_i{}^j{}_k  \omega^k_T,
\end{eqnarray}
\end{subequations}
where we have decomposed $\xi^\mu$ and $\omega^i$ into the scalars $\xi^0$, $\xi$, $\omega$ and the transverse vectors $\xi_T^i$ and $\omega_T^i$ 
($\partial_i \xi_T^i=\partial_i \omega_T^i=0$). Comparison of equations (\ref{general-pert}) and (\ref{general-transformation}) then shows that, by performing an appropriately chosen rotation together with a general coordinate transformation, one can set for instance $F_i =W_i=0$ and $C= D= E  = 0= 0$. Thus, we are eventually left with only four scalars ($\phi, B, \psi$ and $\sigma$), two vectors ($S_i$ and $T_i$) and one tensor ($h_{ij}$). This is the same number of degrees of freedom one obtains in the metric formulation of the theory, after completely fixing the gauge. 

\subsection{Tensor Sector}

As we mention above,  in the free theory,  scalars, vectors and tensors decouple from each other.  Let us therefore start by considering the tensor sector, which is described by the quadratic  Lagrangian
\be
\mathcal{L}_t = \fr{1}{4} \l\{ \l[ M_G^2 - \l(\bar{\beta}_4+\bar{\beta}_5 \r) \Lambda^2 \r] \dot{h}_{ij} \dot{h}_{ij} -  \l[ M_G^2 - \bar{\beta}_4 \Lambda^2 \r] \d_k h_{ij} \d_k h_{ij}\r\},
\ee
from which we can immediately read off the residue and the speed of sound of  the tensor modes,
\be \la{tensor-results}
Z_t^{-1} = \fr{M_G^2 - \l(\bar{\beta}_4+\bar{\beta}_5\r) \Lambda^2 }{2}, \quad\quad  c_t^2 =  \fr{M_G^2 - \bar{\beta}_4 \Lambda^2}{M_G^2 - (\bar{\beta}_4+\bar{\beta}_5)\Lambda^2 } .
\ee
Once again,  $\bar{\beta}_4$ and $\bar{\beta}_5$ stand for the values of the couplings at the minimum of the potential, and a similar notation applies in what follows to the other couplings too. The tensor sector is ghost free provided $(\bar{\beta}_4 +\bar{\beta}_5)\ll(M_G /\Lambda)^{2}$. We should also impose $\bar{\beta}_4 \ll(M_G /\Lambda)^{2}$ in order to ensure classical stability. The results (\ref{tensor-results}) agree with the ones of aether models with parameters given by equation (\ref{mapping}), and they also reduce to the ones found by Gripaios \cite{Gripaios:2004ms} in the limit where $\Lambda \ll M_G$.

\subsection{Vector Sector}
The Lagrangian for the vector modes is only slightly more complicated, and  reads
\begin{eqnarray}\label{eq:vector vector}
\mathcal{L}_v &=&\frac{1}{2}\left\{\left[M_G^2-\bar{\beta}_4 \Lambda^{2}\right]\partial_i (T_j +S_j)\partial_i (T_j +S_j)+2(\bar{\alpha_1}-\bar{\alpha})\Lambda^{2}\,\dot{T}_i \dot{T}_i+ \right.\\
&& \qquad  +(\bar{\beta}_5+2\bar{\alpha})\Lambda^{2}\,\partial_{i}T_j \partial_{i}T_j -\bar{\beta}_5\Lambda^{2}\,\partial_{i}S_j \partial_{i}S_j  \Big\}.\nonumber
\end{eqnarray}
The field $S_i$ only appears in the Lagrangian density through the combination $\partial_i S_j$ and does not propagate. Its equation of motion can be easily solved to get
\be
\left[M_G^2 - (\bar{\beta}_4+\bar{\beta}_5)\Lambda^2 \right]S_i=-\left[M_G^2 - \bar{\beta}_4\Lambda^2 \right]T_i,
\ee
which, when substituted back in (\ref{eq:vector vector}) gives
\be
\mathcal{L}_v = (\bar{\alpha_1}-\bar{\alpha})\Lambda^{2}\,\dot{T}_i \dot{T}_i + \l( \bar{\alpha} -  \fr{ \bar{\beta}_5^2 \Lambda^2}{2 \left[M_G^2 - (\bar{\beta}_4+\bar{\beta}_5)\Lambda^2 \right]}\r)\partial_i T_j \partial_i T_j .
\ee
Therefore, only two massless vector modes propagate, with residue and a speed of sound given by
\be
Z_v^{-1} =2(\bar{\alpha_1}-\bar{\alpha})\Lambda^{2}, \quad \quad c_v^2 = \fr{1}{\bar{\alpha}-\bar{\alpha}_1} \l( \bar{\alpha} -  \fr{ \bar{\beta}_5^2 \Lambda^2}{2 \left[M_G^2 - (\bar{\beta}_4+\bar{\beta}_5)\Lambda^2 \right]}\r). 
\ee
In empty space, the vector sector of general relativity is non-dynamical. However, the breakdown of Lorentz invariance gives dynamics to this sector, even in the absence of matter fields. Of course, these two vector modes correspond to two of the Goldstone bosons of the spontaneously broken phase.
They are well behaved in the limit $\Lambda \ll M_G$ provided $  \l( \bar{\alpha}_1 - \bar{\alpha} \r) > 0$ and $\bar{\alpha}<0$. Notice that this result does not agree with  \cite{Gripaios:2004ms}, though it does agree with the result found in aether theories \cite{Jacobson:2008aj}, upon the identification in equations  (\ref{mapping}).

\subsection{Scalar Sector}

Let us finally consider the scalar sector, which now contains both massive and massless fields. To quadratic order in the perturbations, its Lagrangian density is given by 
\begin{eqnarray}
\mathcal{L}_s &=& \frac{1}{2} \Big\{  2 ( M_G^2 - \bar{\beta}_4 \Lambda^2 ) ( \partial_i \psi \partial_i \psi - 2 \partial_i \phi \partial_i \psi ) + \bar{\beta} \Lambda^2 (\Delta B)^2   \nonumber  \\ 
&& \qquad  + (3\bar{\beta} \Lambda^2 +2\bar{\beta}_{5} \Lambda^2 +2 \bar{\beta}_4 \Lambda^2 -2M_G^2) (3 \dot{\psi}^2 + 2 \dot{\psi} \Delta B ) \nonumber \\
&& \qquad + (\bar{\alpha}_{1} - \bar{\alpha})  \Lambda^2 \partial_i \phi \, \partial_i \phi
+(\bar{\beta}-\bar{\alpha}_{1}-\bar{\alpha}_{2}-\bar{\alpha}_{3}+\bar{\gamma})  \dot{\sigma}^{2} - (\bar{\alpha}-\bar{\alpha}_{2})  \partial_{i}\sigma\partial_{i}\sigma+ \nonumber\\
&& \qquad - 2 \bar{V}'' \Lambda^{2}\sigma^{2}+
[(- 4\bar{\beta}_{4}+ 4\bar{\beta}_{4}' -2 \bar{\beta}_{5}+2\bar{\beta}_{5}'+\bar{\alpha}_{3}-2\bar{\beta})\dot{\sigma}+2\bar{\delta}_1 \Lambda\sigma](\Delta B+3\dot{\psi})+ \nonumber \\
&& \qquad   -\partial_{i}\sigma\partial_{i}[( 4\bar{\beta}_{4}- 4\bar{\beta}_{4}' +2 \bar{\beta}_{5}-2\bar{\beta}_{5}'+2\bar{\alpha})\phi -8(\bar{\beta}_{4}-\bar{\beta}_{4}')\psi]\Big\}. \label{scalar Lagrangian}
\end{eqnarray}
The scalars $\phi$ and $B$ only appear in the Lagrangian trough the combinations $\partial_i \phi$ and $\Delta B$, so they can be easily eliminated by solving their classical equations of motion.  At this point, it is more convenient to switch to Fourier space, and write the action for the two remaining scalars in the form
\begin{equation}\label{eq:timelike L}
S_s= - \frac{1}{2}\int d^4k\,
X^\dag  D X ,\quad\textrm{with} \qquad X \equiv \begin{pmatrix} \sigma(k) \\ \psi(k) \end{pmatrix}
\end{equation}
and
\begin{equation}
D \equiv  \begin{pmatrix} \;\;a_{1} \omega^2 + a_{2} k^2 + a_3 \Lambda^2 \;\;&  a_4 \omega^{2}+ a_5 k^{2}+ia_6\Lambda\,\omega  \\  a_4\omega^{2}+a_5k^{2}-i a_6\Lambda\,\omega & a_7 \omega^2 + a_8 k^2  \end{pmatrix}.
\end{equation}
Here,  the (dimension two) coefficients  $a_i$ are some complicated functions of the various coupling constants of the model. In particular, $a_3$ and $a_6$ are the only couplings that break the $\mathbb{Z}_2$ symmetry $A^a\to -A^a$. 

The inverse of the matrix $D$ is just the field propagator. In order to find the propagating modes we just have to find the values of $\omega^{2}$ at which its eigenvalues have poles, or, equivalently, the values of $\omega^{2}$ at which the eigenvalues of $D$ have zeros. Requiring that det$(D)$ vanish we thus arrive at the frequencies of the two propagating modes,
\begin{equation} \la{scalardisprel}
\omega_{1}^{2}=m_1^{2}\Lambda^{2} + c_1^{2} k^{2} +\mathcal{O}(k^4/\Lambda^2),\qquad \omega_{2}^{2}= c_2^{2}k^2+\mathcal{O}(k^4/\Lambda^2),
\end{equation}
with
\begin{subequations}
\ba
&& m_1^2 = \frac{a_6^2-a_3a_7}{a_1a_7 - a_4^2},
 \\
&&  c_1^2 = \frac{a_8 (a_3 a_4^2-a_1 a_6^2)+(a_6^2-a_3 a_7) (2 a_4 a_5-a_2 a_7)}{(a_6^2-a_3a_7)(a_1a_7 - a_4^2)},\\
&&c_2^{2}=\frac{a_3 a_8}{a_6^2-a_3a_7}.
\ea
\end{subequations}
In the absence of fine-tuning, the first mode has a mass of   order $\Lambda$ and can be excluded from the low-energy theory. On the other hand, the speed of sound of the massless mode, 
\be \la{sos}
c_{2}^2 = \fr{(2 \bar{V}'' \bar{\beta} + \bar{\delta}_1^{2}) \, \left[2 M_G^2 - (2 \bar{\beta}_4 -\bar{\alpha} + \bar{\alpha}_1 ) \Lambda^2 \right] \left[M_G^2 -\bar{\beta}_4  \Lambda^2\right]}{(\bar{\alpha}-\bar{\alpha}_1)\left[M_G^2 -(\bar{\beta}_4 -\bar{\beta}_5)  \Lambda^2\right]\left[2 \bar{V}'' \left(2M_G^2-(2\bar{\beta}_4 +2\bar{\beta}_5 + 3\bar{\beta}) \Lambda^2\right)-3 \,\bar{\delta}_1^2 \Lambda^2 \right]},
\ee
coincides with the speed of sound of the scalar mode in aether theories \cite{Jacobson:2008aj}, after substitution of equations  (\ref{mapping}).  Note that the terms $\mathcal{O}(k^4/\Lambda^2)$ in equation (\ref{scalardisprel}) cannot be trusted since our starting point was an effective action in which all the terms with more than two derivatives were excluded. 

As in the vector sector,
in the absence of matter fields the scalar sector of general relativity is non-dynamical.
But again, the breakdown of Lorentz invariance gives dynamics to this sector. This captures of course the existence of a Goldstone boson in the scalar sector of the theory, which, together with the two massless modes we found in the vector sector, play the role of the three Goldstone bosons associated with the broken boost generators.

The residues of the scalar modes can be determined using the general result \cite{ArmendarizPicon:2009ai}
\begin{equation} \la{resid}
\fr{1}{Z_{1,2}} = -  \fr{1}{\mbox{tr}(D)} \l.\fr{\d}{\d \omega^2} \det(D) \r|_{\omega^2 = \omega^2_{1,2}},
\end{equation}
which, in our case, yields
\begin{equation}
Z_1^{-1} = \frac{a_6^2(a_1+a_7)-a_3(a_4^2+a_7^2)}{(a_1a_7 - a_4^2)(a_6^2-a_3a_7)}+\mathcal{O}(k^4/\Lambda^2), \qquad  Z_2^{-1} = \frac{a_3}{a_3 a_7-a_6^2}+\mathcal{O}(k^4/\Lambda^2).
\end{equation}
Like for the speed of sound, the residue of the massless mode 
\begin{equation}
	Z_2^{-1} = \frac{2\left[ M_G^2 - (\bar{\beta}_4+ \bar{\beta}_5) \Lambda^2\right] 
	\left[ 3\bar{\delta}_1^2\Lambda^2-2\bar{V}''(2M_G^2-(2\bar{\beta}_4+2\bar{\beta}_5+3\bar{\beta})\Lambda^2) \right]}
	{(\bar{\delta}_1^2+2\bar{V}''\bar{\beta})\Lambda^2} +\mathcal{O}(k^4/\Lambda^2)
\end{equation}
agrees with that obtained in aether theories \cite{Jacobson:2008aj}, upon  the identification (\ref{mapping}). Once again, the terms $\mathcal{O}(k^4/\Lambda^2)$ in the residues are out of the reach of validity of the effective theory we wrote down. 

To conclude, it is interesting to point out that none of the results concerning the massless modes  depend on $\alpha_2$, $\alpha_3$, $\gamma$, nor on the derivatives of $\beta_4$ and $\beta_5$. A brute-force approach like the one we just followed makes this look like the result of accidental cancellations. Notice for instance that  in fact the free scalar Lagrangian (\ref{scalar Lagrangian}) does depend on $\alpha_2$, $\alpha_3$, $\gamma$, as well as on the derivatives of $\beta_4$ and $\beta_5$.  The low-energy effective action (\ref{eq:integrated out}) on the other hand  makes this manifest from the very beginning. 

\subsection{The field $\sigma$}

We obtained the low energy effective Lagrangian  (\ref{eq:integrated out}) by integrating out the field $\sigma$. In that context, we claimed that this procedure was justified because that the matrix element of $\sigma$ between the vacuum and a state with one massless particle vanishes in the low-momentum limit (see equation (\ref{matrix-limit})). We are now in a position to prove this result.

As we have seen above, the scalar spectrum consists of a massive field $s_1$ and a massless field $s_2$. We can thus express the field $\sigma$ as a linear combination of the two canonically normalized fields,  
\begin{equation}
	\sigma = \kappa_1 s_1 + \kappa_2 s_2,
\end{equation}
in which $\kappa_1$ and $\kappa_2$ are momentum-dependent coefficients. Therefore, using the reduction formula, the matrix element for emission of a massless excitation in equation (\ref{matrix-limit}) can be written as 
\begin{align}
\langle m = 0, p | \sigma (p') | 0 \rangle &= \lim_{\omega\to \omega_2} i \, (\omega_2^2 - \omega^2) \langle s_2(p) \sigma(p')  \rangle_T= \nonumber  \\ 
	&=  i  \kappa_2 \lim_{\omega\to \omega_2} (\omega_2^2 - \omega^2) \langle s_2(p) s_2(p')  \rangle_T  =  \delta(p+p') \, \kappa_2 ,
\end{align}
where $p = (\omega, k)$, the energy $\omega_2$ was defined in equation (\ref{scalardisprel}), and  $\langle f(p) g(p') \rangle_T$ is the Fourier transform of the corresponding Green's function. The value of  $\kappa_2 $ can be readily calculated by noting that 
\ba
 - i \delta(p+p') D_{\sigma\sigma}^{-1} (p)=  \langle \sigma(p) \sigma(p') \rangle_T &=& \kappa_1^2  \langle s_1(p) s_1 (p')\rangle_T + \kappa_2^2 \langle s_2 (p) s_2 (p') \rangle_T \\
&=& \delta(p+p') \l( \frac{i \kappa_1^2}{\omega^2-\omega_1^2} + \frac{i \kappa_2^2}{\omega^2-\omega_2^2} \r).
\ea
Hence, 
\be
\kappa_2^2 = \lim_{\omega \to \omega_2} (\omega_2^2 - \omega^2) \, D_{\sigma\sigma}^{-1} =  \frac{a_6^2 \, a_8}{(a_3 a_7 - a_6^2)^2} \frac{k^2}{\Lambda^2} + \mathcal{O}( k^4 /\Lambda^4),
\ee
which clearly shows that $\kappa_2$ vanishes in the low-momentum limit.

\end{document}